\documentclass{aa}  

\usepackage{graphicx}
\usepackage[dvipsnames]{xcolor}
\usepackage{caption}
\usepackage{multirow}
\usepackage{txfonts}
\usepackage[colorlinks,allcolors=blue]{hyperref}

\newcommand{\figsizeee}{0.49}
\newcommand{\figsize}{0.315}
\newcommand{\figsizee}{0.32}

\newcommand{\xmm}{{\it XMM-Newton}\xspace}
\newcommand{\nustar}{{\it NuSTAR}\xspace}
\newcommand{\suzaku}{{\it Suzaku}\xspace}
\newcommand{\swift}{{\it Swift}\xspace}
\newcommand{\chandra}{{\it Chandra}\xspace}
\newcommand{\gaia}{{\it Gaia}\xspace}
\newcommand{\HeII}{\ion{He}{ii}\xspace}

\usepackage{threeparttable}
\usepackage{lscape}

\usepackage{txfonts}
\begin{document}

   \title{Periodicity from X-ray sources within the inner Galactic disk}

   \author{Samaresh Mondal\inst{1},
          Gabriele Ponti\inst{1,2},
          Tong Bao\inst{3},
          Frank Haberl\inst{2},
          Sergio Campana\inst{1},
          Charles J. Hailey\inst{4},
          Shifra Mandel\inst{4},
          Sandro Mereghetti\inst{5}
          Kaya Mori\inst{4},
          Mark R. Morris\inst{6},
          Nanda Rea\inst{7,8}, and 
          Lara Sidoli\inst{5}
          }

   \institute{$^1$INAF -- Osservatorio Astronomico di Brera, Via E. Bianchi 46, 23807 Merate (LC), Italy
             \email{samaresh.mondal@inaf.it}\\
             $^2$Max-Planck-Institut f{\"u}r extraterrestrische Physik, Gie\ss enbachstra\ss e 1, 85748, Garching, Germany\\
             $^3$School of Astronomy and Space Science, Nanjing University, Nanjing 210046, China\\
             $^4$Columbia Astrophysics Laboratory, Columbia University, Columbia, NY, 10027, USA\\
             $^5$INAF -- Istituto di Astrofisica Spaziale e Fisica Cosmica, via A. Corti 12, 20133 Milano, Italy\\
             $^6$Department of Physics and Astronomy, University of California, Los Angeles, CA, 90095-1547, USA\\
             $^7$Institute of Space Sciences (ICE, CSIC), Campus UAB, Carrer de Can Magrans s/n, E-08193 Barcelona, Spain\\
             $^8$Institut d'Estudis Espacials de Catalunya (IEEC), Carrer Gran Capit\`a 2--4, 08034 Barcelona, Spain
             }

   \date{Received XXX; accepted YYY}
   \authorrunning{Mondal et al.}
   \titlerunning{Periodic X-ray sources in the Galactic Center and disk}

 
  \abstract
   {}
   {For many years it had been claimed that the Galactic ridge X-ray emission at the Galactic Center (GC) is truly diffuse in nature. However, with the advancement of modern X-ray satellites, it has been found that most of the diffuse emission actually comprises thousands of previously unresolved X-ray point sources. Furthermore, many studies suggest that a vast majority of these X-ray point sources are magnetic cataclysmic variables (CVs) and active binaries. One unambiguous way to identify these magnetic CVs and other sources is by detecting their X-ray periodicity. Therefore, we systematically searched for periodic X-ray sources in the inner Galactic disk, including the GC region.}
   {We used data from our ongoing \xmm Heritage Survey of the inner Galactic disk ($350\degr\lesssim l\lesssim+7\degr$ and $-1\degr\lesssim b\lesssim +1\degr$) plus archival \xmm observations of the GC. We computed the Lomb-Scargle periodogram for the soft (0.2--2 keV), hard (2--10 keV), and total (0.2--10 keV) band light curves to search for periodicities. Furthermore, we modeled the power spectrum using a power-law model to simulate 1000 artificial light curves and estimate the detection significance of the periodicity. We fitted the energy spectra of the sources using a simple power-law model plus three Gaussians, at 6.4, 6.7, and 6.9 keV, for the iron $K$ emission complex.}
   {We detected periodicity in 26 sources. For 14 of them, this is the first discovery of periodicity. For the other 12 sources, we found periods similar to those already known, indicating no significant period evolution. The intermediate polar (IP) type sources display relatively hard spectra compared to polars. We also searched for the \gaia counterparts of the periodic sources to estimate their distances using the \gaia parallax. We found a likely \gaia counterpart for seven sources.}
   {Based on the periodicity, hardness ratio, and the equivalent width of Fe $K$ line emission, we have classified the sources into four categories: IPs, polars, neutron star X-ray binaries, and unknown. Of the 14 sources for which we detect the periodicity for the first time, four are likely IPs, five are likely polars, two are neutron star X-ray binaries, and three are of an unknown nature.}


   \maketitle
   
%
\section{Introduction}
In order to understand the star formation history of our Galaxy, it is important to know the number of stars that ended their main-sequence life long ago. Compact remnants of dead stars, such as black holes, neutron stars (NSs), and white dwarfs (WDs), are commonly found in binary systems and are visible in X-rays. Accreting WD binaries are the most common type of remnant in our Galaxy as WDs are the end product of intermediate- and low-mass stars. Many of these low-mass stars are born in binary systems with small separations that go through one or more mass transfer phase, leading to the formation of cataclysmic variables (CVs). More than a thousand CVs have been found in the solar neighborhood \citep{downes2001,ritter2003}.  CVs are categorized as magnetic or nonmagnetic \citep[see][for a review]{copper1990,petterson1994,mukai2017}. Magnetic CVs are primarily categorized into two subtypes: polar and intermediate polar (IP). Polars have a very strong magnetic field ($>10$ MG), which synchronizes the spin and orbital motion (i.e., $P_{\rm spin}=P_{\rm orb}$). The high magnetic field in polars is confirmed by the observation of strong optical polarization and the measurement of cyclotron humps \citep{warner2003}. In polars, the accretion directly follows the magnetic field lines from the L1 point, and no accretion disk is formed. IPs have a relatively weak surface magnetic field of $1-10$ MG; therefore, they have less synchronization, and an accretion disk is created. In these systems, the material leaving the L1 point forms an accretion disk until the magnetic pressure becomes equal to the ram pressure of the accreting gas. The X-ray emission from CVs originates from close to the magnetic pole. The accreting material follows the magnetic field lines, and as it approaches the WD surface, the radial in-fall velocity reaches supersonic speeds of  3000-10000 km s$^{-1}$. A shock front appears above the WD surface, and the in-falling gas releases its energy in the shock, resulting in hard X-ray photons \citep{aizu1973,saxton2005}. 

Early observations of the Galactic Center (GC) revealed a diffuse X-ray emission \citep{worrall1982,warwick1985,yamauchi1996} called Galactic ridge emission. For many years a central point of debate has been whether the Galactic ridge emission is truly diffuse
or composed of emission from many unresolved X-ray point sources. The advent of modern X-ray satellites such as \xmm and \chandra opened up the possibility of detecting very faint X-ray sources in crowded regions such as the inner GC. This is not possible in the optical waveband due to the high extinction toward the GC. A deep \chandra observation of the inner Galactic bulge has demonstrated that more than 80\% of the Galactic ridge emission is produced by CVs and coronally active stars \citep{wang2002,revnivtsev2009,muno2003,muno2009,zhu2018}. Although 
 this strongly indicates that a large fraction of the X-ray sources observed toward the GC are magnetic CVs, the physical nature of CVs in the GC remains unclear. Moreover, it was suggested that, based on the hard power-law-type spectral shape and the emission of Fe $K$ complex lines, the majority are IPs  \citep{muno2004}
 
 The Galactic ridge X-ray emission displays a copious amount of lines from ionized iron at 6.7 and 6.9 keV. Some studies that compared the stellar mass distribution with the Fe XXV (6.7 keV) line intensity map suggest the presence of truly diffuse hard X-ray emission  \citep{uchiyama2011a,nishiyama2013,yasui2015}. However, a recent study by our group found that this diffuse hard emission in the GC can be explained if one assumes that the GC stellar population has iron abundances $\sim1.9$ times higher than those in the Galactic bar/bulge \citep{anastasopoulou2023}. Furthermore, the 20--40 keV emission from the GC observed by \nustar is best explained by two-temperature plasma models with $kT_1\sim1$ keV and $kT_2\sim7.5$ keV. The $\sim1$ keV temperature component is attributed to emission from supernovae heating the interstellar medium, coronally active stars, and nonmagnetic WDs \citep{revnivtsev2009}. The $\sim7.5$ keV temperature component is thought to be produced by emission from resolved and unresolved accreting IPs \citep{perez2015}. An additional component with a higher plasma temperature, $kT\sim35$ keV \citep{hailey2016}, was recently measured. In addition, \citet{muno2003a} reported the discovery of eight periodic sources in a $17'\times17'$ field of the GC. Their periods range from 300 s to 4.5 hr. All these sources exhibit hard power-law-type spectral shapes (with photon index $\Gamma\sim0$) and 6.7 keV iron-line emission. These properties are consistent with magnetically accreting magnetic CVs.

We are in the process of performing an X-ray scan of the inner Galactic disk using \xmm (\citealt{jansen2001}; PI: G. Ponti). The main aim of this survey is to constrain the flow of hot baryons that feed large-scale energetic events such as the Galactic chimneys \citep{ponti2019,ponti2021}, the \emph{Fermi} bubbles \citep{su2010}, and the eROSITA bubbles \citep{predehl2020}. In early 2021, while performing this survey, we detected an X-ray source with periodic modulation at 432 s \citep{mondal2022}. The source was previously observed by \suzaku in 2008 \citep{uchiyama2011b} and classified as an IP. Furthermore, while examining \xmm archival observations, we discovered periodicity in two other sources within 1\fdg5 of the GC \citep{mondal2023}. These two sources are also classified as IPs based on the detected spin period and detection of an iron emission complex in the spectra. Therefore, we took a systematic approach to hunt for such periodic X-ray sources that might help us classify them. In this paper we report the discoveries obtained from a periodicity search using \xmm observations of the Galactic disk and the GC.

\section{Observations and data reduction}
\begin{figure*}
    \centering
    \includegraphics[width=0.99\textwidth]{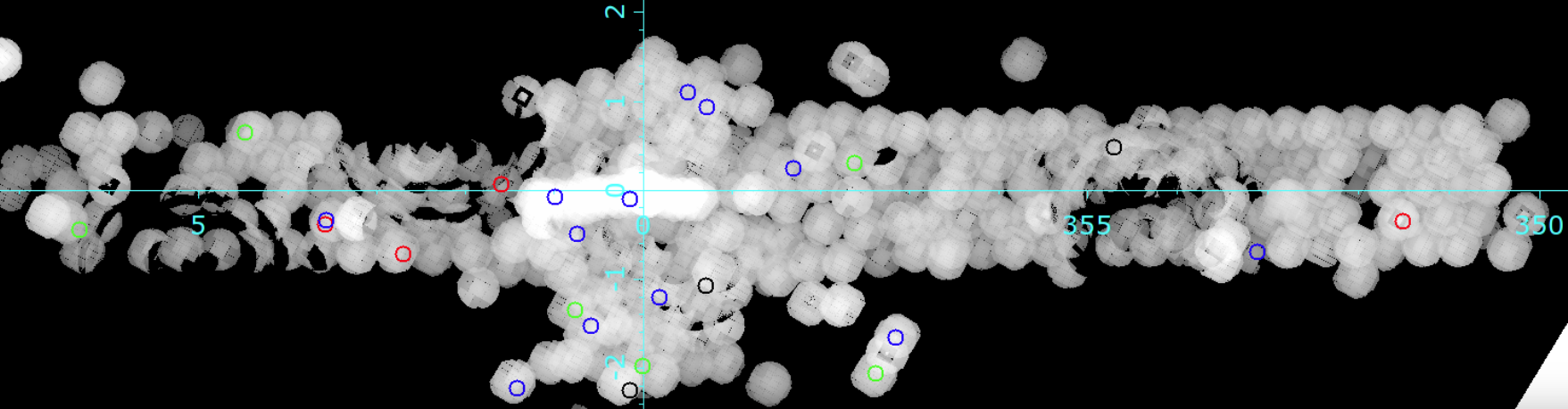}
    \caption{Mosaic of the exposure maps created using the ongoing \xmm observations of the Galactic disk plus archival observations of the GC. The small red, blue, and green circles show the positions of confirmed or likely NSs, IPs, and polars, respectively. The black circles indicate the unclassified sources.}
    \label{fig:pos}
\end{figure*}

 We have almost completed one side of the Galactic disk extending from $l\geq350\degr$ to $l\leq+1.5\degr$ (see Fig. \ref{fig:pos}). The survey has an exposure of 20 ks per tile and is expected to cover the Galactic disk region in the range $350\degr<l<+7\degr$ and $-1\degr<b<+1\degr$. During this campaign, we detected thousands of X-ray point sources of various types. A forthcoming paper will present a sample study of the X-ray point sources. Here we are focusing on X-ray sources that show periodic modulations. While doing this analysis, we considered including the GC region for positional comparison of the sources located in the disk and GC \citep{ponti2013,ponti2015}. For the GC, we used the \xmm archival observations. In total, we analyzed 444 \xmm observations, including our Galactic disk scanning observations plus the archival observations of the GC. The observation data files were processed using the \xmm Science Analysis System (SAS, v19.0.0)\footnote{https://www.cosmos.esa.int/web/xmm-newton/sas}. We used the task \texttt{evselect} to construct a high energy background light curve (energy between 10 and 12 keV for EPIC-pn and above 10 keV for EPIC-MOS1 and MOS2) by selecting only PATTERN==0. The background light curve was used to filter high background flaring activity and create good time intervals. We used the SAS task \texttt{emldetect} for point source detection and source list creation. For each individual detector, EPIC-pn, MOS1, or MOS2 \citep{struder2001,turner2001}, the source detection was performed in five energy bands: 0.2--0.5 keV, 0.5--1 keV, 1--2 keV, 2--4 keV, and 4--12 keV for a given single observation. The source detection algorithm separately provides net counts and maximum likelihood values for the five energy bands and three detectors: EPIC-pn, MOS1, and MOS2. The source detection tool also provides the keyword \texttt{EXT} value that indicates whether the emission is from a point-like or extended source. We chose \texttt{EXT} $=0$ to select the point sources only. The total number of point sources detected in our survey is $\sim50000$. Then, we applied a filter in which only sources with a total number of counts (EPIC-pn+MOS1+MOS2) higher than 200 were chosen. This resulted in 2500 point sources for which we extracted the light curves using the SAS task \texttt{evselect} after applying the Solar System barycenter correction to the event files using the SAS task \texttt{barycen}. We only selected events with PATTERN$\le$4 and PATTERN$\le$12 for EPIC-pn and the MOS1 and MOS2 detectors, respectively. We chose a circular region of 20\arcsec\ radius for the source products extraction. The background products were extracted from an annular region centered on the source position with inner and outer radii of 25\arcsec\ and 30\arcsec, respectively. The spectra were binned to have a minimum of 20 counts in each energy bin. Many fields of the GC have been observed more than once. If a source has been observed multiple times, we searched for pulsations in each observation individually.


\section{Results}

\subsection{Period search}
\label{subsec:search}
The \xmm observations suffer from gaps due to the filtering of high background flaring activities. As the \xmm observations suffer from gaps, we used the Lomb-Scargle periodogram \citep{lomb1976,scargle1982}, which is well known for detecting periodic signals in unevenly sampled time series data. We computed the false alarm probability to estimate the statistical significance of the periodogram peaks. The false alarm probability obtained is based on the analytical approximation proposed by \citet{baluev2008}, which employs extreme value statistics to compute an upper bound of the false alarm probability (or a lower limit of the significance of the detected periodicity). For our timing analysis, we used the {\sc PYTHON}-based \texttt{astropy} \citep{astropy_collaboration2013,astropy_collaboration2018,astropy_collaboration2022} package's time-series module\footnote{https://docs.astropy.org/en/stable/timeseries/index.html}. 

We extracted the light curves in three different bands (0.2--2, 2--10, and 0.2--10 keV) for the three EPIC detectors (pn, MOS1, and MOS2). The light curves were extracted with a time bin of 74 ms for the EPIC-pn and 3 s for the MOS1 and MOS2 detectors in full frame mode of observation. The Lomb-Scargle periodogram was computed for all nine light curves of each source, and a periodicity search was conducted. The EPIC-pn detector has a frame time of 74 ms, which allowed us to probe a maximum frequency of $\sim6$ Hz, whereas in the case of the MOS1 and MOS2 detectors, we were able to probe a maximum frequency of $\sim0.16$ Hz. We imposed the criterion that the periodicity detected at a frequency below 0.16 Hz should be present in all three detectors. To search for periodicity at a higher frequency within the range 0.16--6 Hz, we used only the data from EPIC-pn. 
We have detected periodicity at a significance above $3\sigma$ in 23 sources. Possible periodicities with significance between 2 and 3$\sigma$ were found in another three sources\footnote{We also detected a few sources with detection significance of the pulsation just above the 1$\sigma$ confidence level. We did not list these sources in Table \ref{table:list_tab}; one such example is the transient Galactic bulge IP XMMU J175035.2--293557 \citep{hofmann2018}.}. 
Figures \ref{fig:psd1} and \ref{fig:psd2} show the Lomb-Scargle periodograms of the 26 sources, with the horizontal lines indicating the detection significance levels.

Figure \ref{fig:pos} shows the mosaic of the exposure maps created from our ongoing \xmm observations of the Galactic disk plus the \xmm archival observations of the GC. The small circles indicate the positions of the periodic sources. We have completed the survey of one side of the Galactic disk extending to $\sim350^{\circ}$; however, most pulsators are concentrated near the GC. 

\begin{figure}
    \centering
    \includegraphics[width=\figsizeee\textwidth]{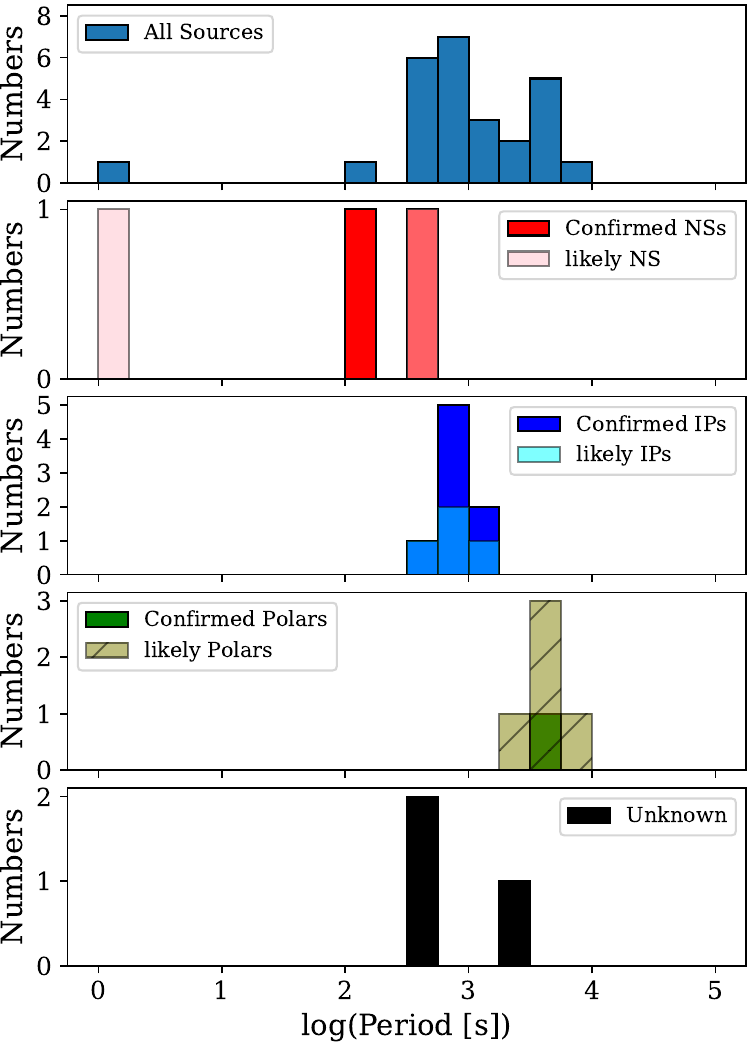}
    \caption{Distribution of log(period) for various source types.}
    \label{fig:hist_period}
\end{figure}

\begin{figure}
    \centering
    \includegraphics[width=\figsizeee\textwidth]{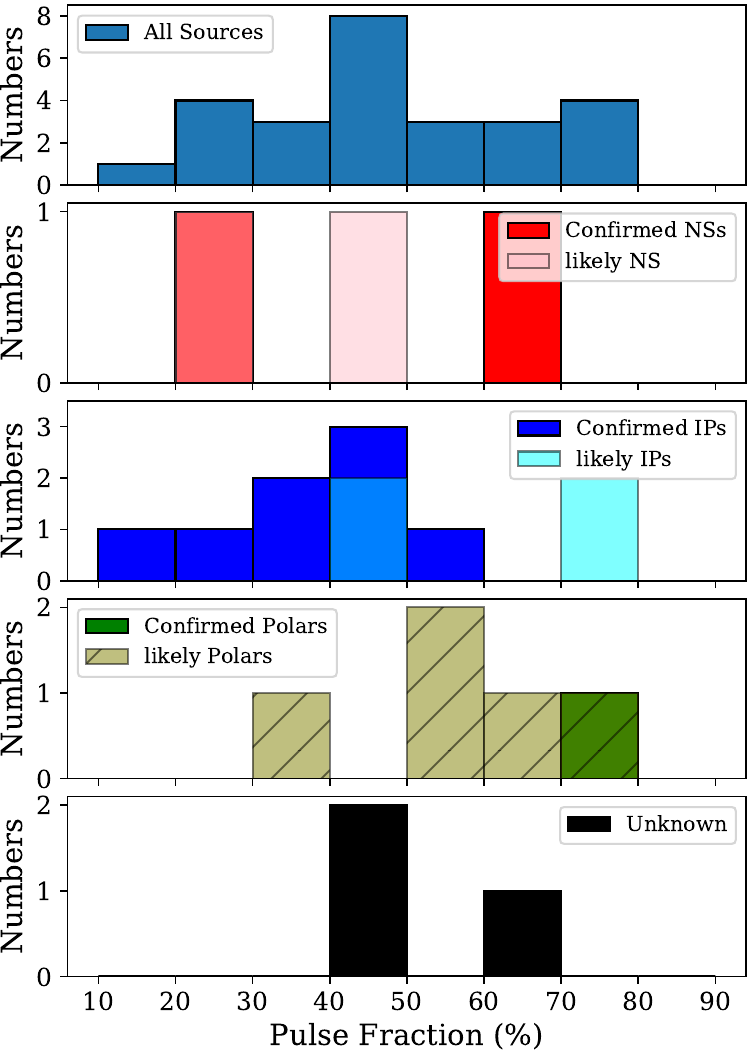}
    \caption{Distribution of the 2-10 keV pulse fraction in percent for different source types.}
    \label{fig:hist_pf}
\end{figure}

Table \ref{table:list_tab} shows the details of the X-ray properties of the periodic sources. The period column shows the pulsation period obtained in our analysis and compares it with the previously reported period. The pulse fraction is computed in 2--10 keV bands. The name of the sources is taken from the 4XMM catalog except for the sources XMMU J173029.8--330920, XMMU J175441.9--265919, and XMMU J180140.3--23422, which were not listed in the 4XMM \citep{webb2020} archive as these sources were first detected in our campaign. The X-ray position and $1\sigma$ positional error of the sources are taken from the 4XMM catalog. We also list the source type based on previous studies, and for sources that were not classified before, we give a tentative classification based on the X-ray period, hardness ratio (HR) values, and spectral properties. Figure \ref{fig:hist_period} shows the distribution of the log of pulse period for various source types. Figure \ref{fig:hist_pf} shows the distribution of pulse fraction for the different categories. The pulse fraction was computed as $\rm PF=\frac{F_{max}-F_{min}}{F_{max}+F_{min}}$, where $\rm F_{max}$ and $\rm F_{min}$ are the maximum and minimum counts in the folded light curves, respectively. For sources with more than one \xmm observation, we report the periodicity from the multiple observations in Tables \ref{table:list_tab1} and \ref{table:list_tab2}.

\subsection{Caveats for the false alarm probability}
Accretion-powered systems such as CVs and X-ray binaries are known to exhibit aperiodic variability over a wide range of timescales. This irregularity, often referred to as "red noise," constitutes a significant aspect of aperiodic variability and has the potential to introduce spurious periodic signals, especially at lower frequencies \citep{1989IBVS.3383....1W}. Consequently, it is essential to assess the likelihood of false detections among these periodic signals found with the Lomb-Scargle periodogram method by using a large simulated dataset \citep{2022MNRAS.509.3504B}.  

Specifically, we employed a power-law model to characterize the source power spectrum, which has the form of
\begin{equation}
P(\nu)=N \nu^{-1}+C_{\rm p}.
\label{eqn:LMXBps}
\end{equation}
In this equation, $N$ represents the normalization factor, and $C_{\rm p}$ accounts for the Poisson noise, which is influenced by the mean photon flux of the source. 

To begin, we estimated the power spectral density (PSD) using the standard periodogram approach with an $\rm [rms/mean]^2$ PSD normalization \citep{2003MNRAS.345.1271V}. However, as mentioned in Section \ref{subsec:search}, some of the light curves suffer from gaps due to background flares. For these cases, we filled the gap with artificial light curves of Poisson noise, assuming the mean flux is consistent with that of the source. Although such processing results in little differences in the described PSDs, for most of the periodic sources here these gaps are fortunately negligible in terms of time (i.e., they take less than 0.5\% of the total exposure time). Only one case exhibits a significant data gap, which takes $\sim 1.4\%$ of the single observation, with ObsID=0783160101. This source (4XMM J174816.9-280750) consistently exhibits the same periodic signal across multiple observations (see Table \ref{table:list_tab2}). Thus, the possible uncertainty of its confidence estimation by the process of filling gaps will not impact the verification of its periodicity.

We fitted the power spectrum of each source with Eq.~\ref{eqn:LMXBps}, using the maximum likelihood function discussed in \citet{2010MNRAS.402..307V} and the Markov chain Monte Carlo approach, employing the Python package \emph{emcee}\footnote{https://emcee.readthedocs.io/en/stable/} \citep{2013PASP..125..306F} to derive the best-fit parameters and their associated uncertainties. 

It turns out that only three of the periodic sources could be adequately described by the power-law model with constrained normalization, implying a potential influence of red noise. For the remaining sources, Poisson noise actually dominates the source power spectrum. Thus, for the source with potential red noise, simulated light curves for this best-ﬁfit power-law model were constructed using the method of \citet{1995A&A...300..707T}, which were resampled and binned to have the same duration, mean count rate, and variance as the observed light curve. As for sources where Poisson noise prevailed, we followed a similar procedure to simulate their light curves, assuming pure Poisson noise. A group of 1000 simulated time series was produced for each source. To evaluate the false alarm level, we computed the maximum Lomb-Scargle power for each simulated light curve. Specifically, we considered the top 0.3\% of the maximum Lomb-Scargle power from the 1000 Lomb-Scargle periodograms, corresponding to the 3$\sigma$ confidence level estimation, and the top 5\% as the threshold corresponding to 2$\sigma$ (approximately 95\%). These simulated thresholds were then overlaid on the Lomb-Scargle periodogram (Figs. \ref{fig:psd1} and \ref{fig:psd2}), and the confidence levels, calculated using Baluev's analysis method \citep{baluev2008}, were compared. It turns out that 23 sources exceed the simulated-based threshold of 3$\sigma$, and 17 of them exceed the 3$\sigma$ threshold of Baluev's method. The deviation between these two is mainly due to that the Baluev method, by design, provides an upper limit to the false alarm probability with little overestimation \citep{baluev2008}.

\subsection{Period and pulse fraction distribution}
The top panel of Fig. \ref{fig:hist_period} shows the period distribution of sources in our sample. The distribution has two peaks at around $\sim800$ s and $\sim4800$ s. The first peak is associated with the population of IPs, and the second peak corresponds to the population of polars.

The spin period of NS and likely NS systems in our sample ranges from 1.36 s to 411.3 s. The spin period of Galactic NS high-mass X-ray binaries (HMXRBs) ranges from a few to thousands of seconds, and the distribution has peaks around $\sim250$ s \citep{Neumann2023}. The red histogram in the top panel of Fig. \ref{fig:hist_period} shows the distribution of the period for NS X-ray binaries in our sample. 

The blue and cyan histogram in the middle panel of Fig. \ref{fig:hist_period} shows the period distribution for the known IPs plus the tentative identification of IPs in our sample. The distribution has a peak of around 607 s. Typically, the spin period of IPs ranges between 30 s and 3000 s, with a peak near 1000 s \citep{scaringi2010}. The middle panel of Fig. \ref{fig:hist_pf} (blue and cyan histogram) shows the distribution of pulse fraction for IPs. The pulse fraction ranges from 10\% to 80\% with a peak near 45\%. One prominent feature in IPs is that the pulse fraction or the modulation depth typically increases with decreasing X-ray energy. This has been thought to be the effect of photoelectric absorption \citep{norton1989a}. The distribution of pulse fraction of IPs covers a wide range of scales and can vary from a few percent up to $\sim$100\% with an average around 24\% \citep{norton1989a,haberl1995}.

In polars, the spin and orbital periods are synchronized, ranging from 3000 s to 30000 s \citep{scaringi2010}. In our sample, the periods of polars vary from 4093 s to 6784 s, with the peak at $\sim4800$ s. The light curves of polars show constant modulation of depth with X-ray energies. The depths are generally higher compared to IPs \citep{norton1989a}. In the middle panel of Fig. \ref{fig:hist_pf}, it is evident that the pulse fraction for polars starts at higher values, around 30\% compared to IPs, and that more polar type sources are found between 50\% and 60\%.

\subsection{Spectral modeling}
We performed time-averaged spectral modeling using the X-ray spectral fitting software {\sc xspec}\footnote{https://heasarc.gsfc.nasa.gov/xanadu/xspec/} \citep{arnaud1996}. We employed $\chi^2$ statics in our model fitting. The spectra were fitted using a simple model composed of a power law and three Gaussian lines (\texttt{tbabs(power-law+g1+g2+g3)}). The Galactic absorption component is represented by \texttt{tbabs} \citep{wilms2000}. For the continuum, we used a simple power-law model, and \texttt{g1}, \texttt{g2}, and \texttt{g3} represent the three Gaussian lines at 6.4, 6.7, and 6.9 keV, respectively, for iron emission complex. While doing the fit, we freeze the line energies at the expected values, and the width of the lines is fixed at zero eV. We jointly fit the spectra of EPIC-pn, MOS1, and MOS2 detectors. While fitting the spectra, we included a constant factor for cross-calibration uncertainty, which is fixed to unity for EPIC-pn and allowed for variation for MOS1 and MOS2. The spectral fitting results are summarized in Table \ref{table:spec_tab}, and Figs. \ref{fig:spec1} and \ref{fig:spec2} show the fitted spectra of the sources.

\begin{figure}
    \centering
    \includegraphics[width=\figsizeee\textwidth]{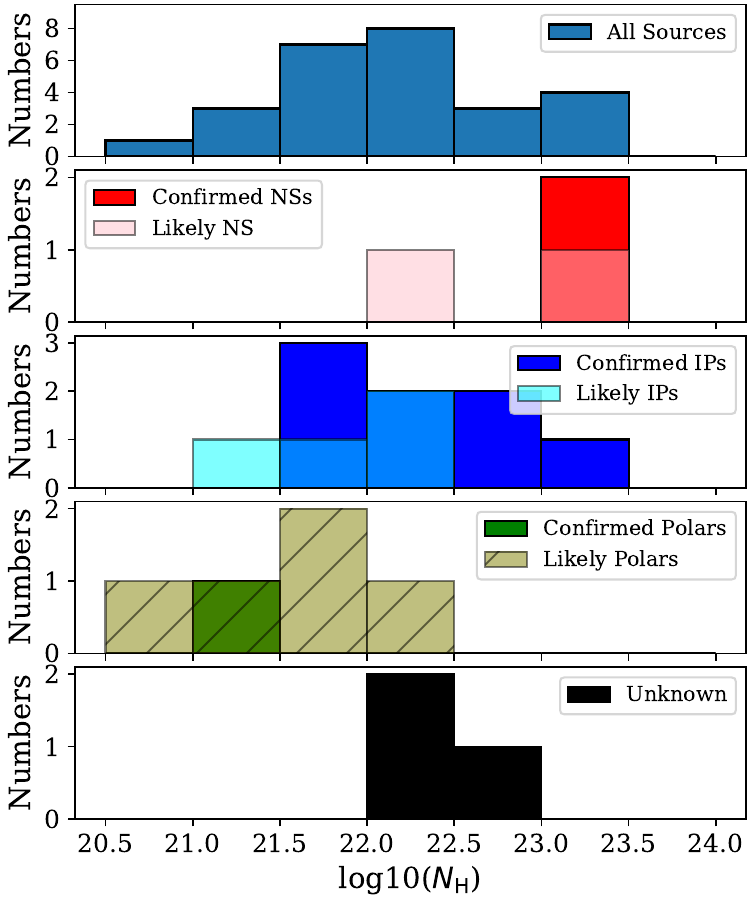}
    \caption{Distribution of the absorption column density, $N_{\rm H}$, for different source types.}
    \label{fig:nh}
\end{figure}

Figure \ref{fig:nh} shows the distribution of absorption column density $N_{\rm H}$ obtained from the X-ray spectral fitting. Overall, the $N_{\rm H}$ distribution has a peak near $10^{22}$ cm$^{-2}$, and more than 50\% of the sources have $N_{\rm H}$ between $10^{21}-3.16\times10^{22}$ cm$^{-2}$. There are three sources with high $N_{\rm H}$>$10^{23}$ cm$^{-2}$. The source 4XMM J175327.8--295716 has $N_{\rm H}=7\times10^{20}$ cm$^{-2}$, the lowest in our sample, which might indicate that this source is the closest to us among our sample or has a soft component that mimics the low $N_{\rm H}$. 

\begin{figure}
    \centering
    \includegraphics[width=\figsizeee\textwidth]{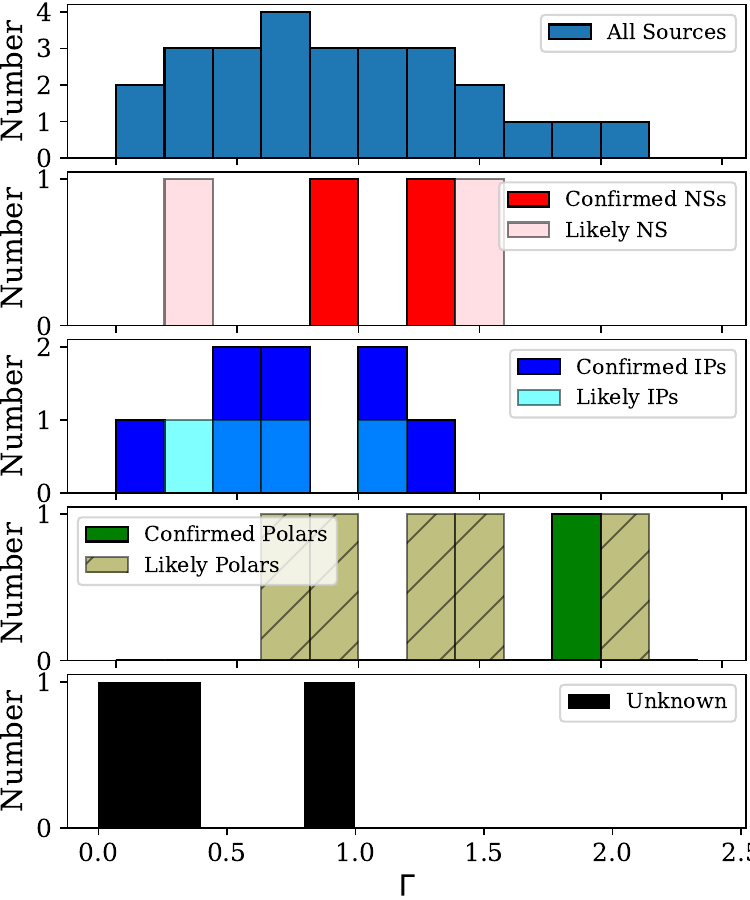}
    \caption{Distribution of the photon index, $\Gamma$, for different source types.}
    \label{fig:gamma}
\end{figure}

Figure \ref{fig:gamma} shows the distribution of photon index $\Gamma$. The distribution has a peak at $\Gamma\sim0.6$. More than 50\% of the sources have a flat spectral shape with $\Gamma$<1. A significant number of sources in our sample have a softer spectrum with $\Gamma$>1.2. We noticed that the majority of sources with high $\Gamma$ values do not show any iron emission complex lines; only two of the seven sources with $\Gamma\geq1.3$, show strong emission lines in the 6--7 keV band.

\subsection{\gaia counterparts}
A correct estimate of the distance to the source is required to derive their luminosity. For this, we searched for counterparts in the \gaia DR3 catalog. For each X-ray source, we computed the \gaia source density by counting the number of \gaia sources within a circle of 1\arcmin\ radius at the source position. The \gaia density of sources is low and varies from 0.01--0.1 arcsec$^{-2}$. We compute the probability of the sources having a spurious association by multiplying the \gaia source density with the area associated with the \xmm positional error. Table \ref{table:list_tab_optical} lists the sources for which we found a \gaia counterpart within the 3$\sigma$ positional uncertainty of \xmm. We found a likely \gaia counterpart for seven \xmm sources. If a counterpart is found, then we use the \gaia source ID to find the distance to the source from \citet{bailer-jones2021}. The distance to the sources for which a \gaia counterpart was found varies from $\sim1.5$ to $\sim5$ kpc and the X-ray luminosity is in the range $5\times10^{32}-6\times10^{33}$ erg s$^{-1}$.

\section{Discussion}
\subsection{Typical properties of different classes of sources}
\label{sec:catagory_class}
We analyzed 444 \xmm observations of the GC and the Galactic disk. We extracted X-ray light curves from nearly 2500 sources and systematically searched for X-ray pulsation. We detected periodicity in 26 sources, 14 of which are reported here for the first time. Many of the GC sources have a luminosity of a few times $10^{32}$ erg s$^{-1}$ \citep{muno2003}, which is comparable to the luminosity typically observed in bright magnetic CVs \citep{verbunt1997,ezuka1999}. NS HMXRBs have much higher luminosity and are detected during their outburst period, reaching luminosities up to $10^{38}$ erg s$^{-1}$. For the majority of the sources we did not find a \gaia counterpart due to the high absorption column density toward the GC and disk. Hence, the X-ray luminosity cannot be derived for a large number of sources and we cannot use this information to classify them. The NS HMXRBs are far less common in our Galaxy than the magnetic CVs. The nature of the Galactic sources is a long-standing question. Identifying the magnetic CVs or NS HMXRBs from X-ray periodicity alone can be difficult, as both types of sources usually display periods in a similar range. The short-period modulation in the X-ray light curve is thought to have originated from the spin period of the magnetically accreting WD or NS. In our sample, the smallest detected period is 1.36 s, and the maximum period detected is around 6784 s. The sample has a median period of 672 s. The pulse fraction of the modulation ranges from 10\% to 80\%. The detected periods are consistent with those of magnetic CVs and NSs in HMXRBs. A sample study of magnetic CVs indicates the median spin period is 6000 s \citep{ritter2003}. There are also a few magnetic CVs with very short spin periods; for example, CTCV J2056--3014 has a spin period of 29.6 s \citep{oliveira2020}, and V1460 Her has a spin period of 38.9 s \citep{ashley2020}. The spin periods of polars are mostly beyond 1 hr, while almost all IPs have WD spin periods lower than 1 hr. In contrast, 85 Galactic HMXRB pulsars (both with Be and OB supergiant companions) have a median (mean) spin period of 187 s ($\sim$970 s), with only four sources showing a period longer than 1 hr \citep{Neumann2023}.

It is evident that the different classes of sources (NS HMXRBs and magnetic CVs)  exhibit a wide range of spin periods. Therefore, from the periodicity alone, it is difficult to understand the nature of the unclassified sources. Below we summarize a scheme to characterize the different classes of periodic X-ray sources utilizing their X-ray spectral, timing properties, and luminosity.
\subsubsection{NS HMXRBs}
\begin{figure}
    \centering
    \includegraphics[width=\figsizeee\textwidth]{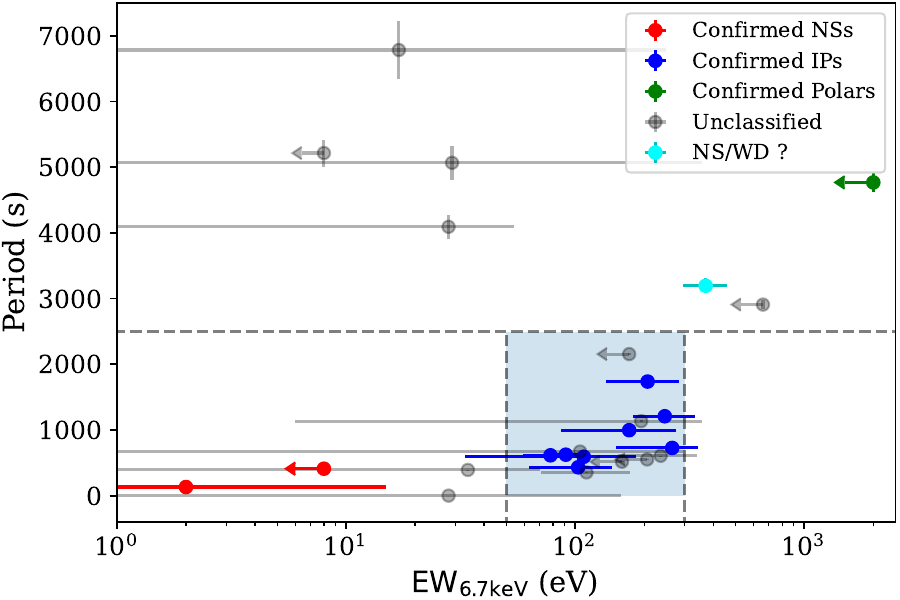}
    \caption{EW of the 6.7 keV line vs. period diagram.}
    \label{fig:period_ew2}
\end{figure}

The NS HMXRBs have properties that are very similar to IPs and they typically have very hard spectra. Figure \ref{fig:period_hr} shows the period versus HR plot for classified and unclassified sources in our sample. The HR is calculated using the net counts in the 2–5 keV and 5–10 keV bands. We did not choose an energy band below 2 keV simply because it would be affected by Galactic absorption. The known NS HMXRBs appear very hard, similar to IPs; however, it is clear from Fig. \ref{fig:period_ew2} that they emit very little 6.7 keV iron line as compared to IPs. In almost all NS HMXRBs, the dominant component of the Fe K emission complex is the neutral 6.4 keV line emission and little to no ionized 6.7 and 6.9 keV line emission. 
This is because HMXRBs are mainly wind-fed systems, so the fluorescent iron line emission from the wind of the companion star is the main spectral feature in their spectra, while the ionized iron emission lines usually come from an accretion disk.
The known NS HMXRBs in our sample -- 4XMM J172511.3–361657 and 4XMM J174906.8–273233 -- show no 6.7 keV emission, with upper limits on their equivalent widths (EWs) of 8 eV and 15 eV, respectively. We define the following criteria for the characterization of the NS HMXRB: (i) $P_{\rm spin}\lesssim$1000 s,  
(ii) HR>-0.2, (iii) $\rm EW_{6.7}$<50 eV,  and
(iv) a typical X-ray luminosity of $10^{33}-10^{37}$ erg s$^{-1}$.

\subsubsection{IPs}
One of the prominent features of IPs is the presence of strong ionized 6.7 keV line emission. In our sample, all the confirmed IPs have a clear detection of a 6.7 keV line, with the lowest EW of the sample being $78^{+34}_{-19}$ for 4XMM J174517.0--321358. \citet{xu2016} studied a sample of bright 17 IPs using \suzaku data. They found that the minimum and mean EW of the 6.7 keV line of the sample is $58^{+10}_{-13}$ and $107\pm17$ eV, respectively. The below criteria can be used to characterize IPs. They typically have 
(i) a spin period $P_{\rm spin}$<2500 s, (ii) an HR>-0.2, 
(iii) a strong 6.7 keV line emission with $\rm EW_{6.7}$>50 eV, and (iv) and an X-ray luminosity in the range $10^{31}-10^{35}$ erg s$^{-1}$ \citep{suleimanov2022}.
\subsubsection{Polars}
The X-ray emission from polars is much softer than that of IPs. The spectra of many polars are dominated by very soft blackbody-like emission from the WD surface \citep{osborne1986,ramsay1993,clayton1994}. However, toward the GC this component is difficult to observe due to the high absorption. In general polars also show a strong 6.7 keV line with an EW anywhere from $50$ eV to $\sim450$ eV \citep{ezuka1999,xu2016}. As a whole, the detection of a 6.7 keV line in polar can be difficult for faint sources as they are much softer than IPs. Polars can be tentatively classified by having softer spectra and periods above 2500 s; however, a secure classification would require the detection of the 6.7 keV line with good quality X-ray spectra and strong circular polarization in the optical band. The polars can be characterized by the following characteristics:  
(i) $P_{\rm spin}=P_{\rm orb}$>2500 s, (ii) HR<-0.2, 
(iii) a strong 6.7 keV line emission with $\rm EW_{6.7}$>50 eV, and 
(iv) an X-ray luminosity below $10^{33}$ erg s$^{-1}$ \citep{suleimanov2022}.

\begin{figure}
    \centering
    \includegraphics[width=\figsizeee\textwidth]{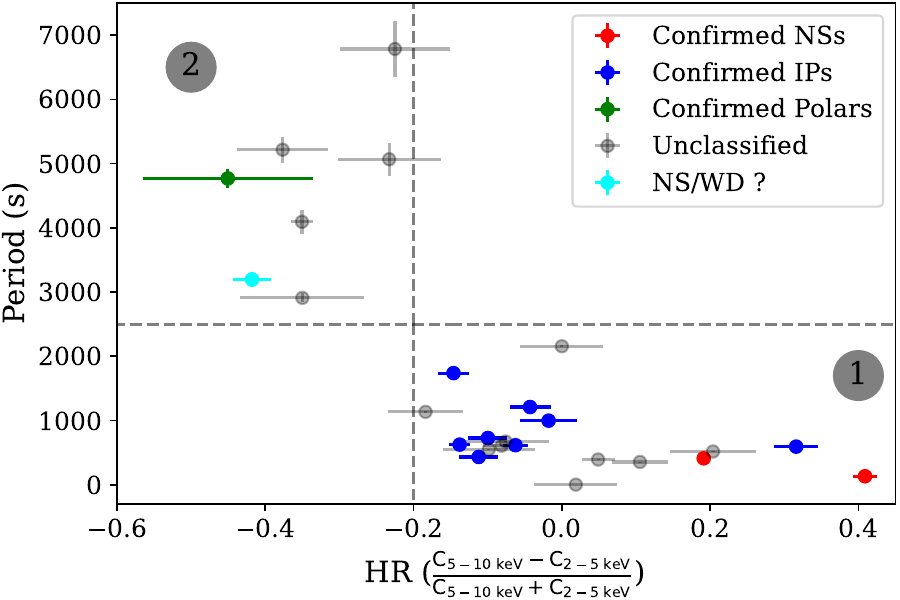}
    \caption{HR vs. period diagram. The HR is calculated using the net counts of two bands: 2--5 keV and 5--10 keV.}
    \label{fig:period_hr}
\end{figure}

\subsection{Known NS HMXRBs}
The source 4XMM J172511.3--361657 was discovered on 9 February 2004 by INTEGRAL and named as IGR J17252--3616 \citep{walter2004}. \xmm observed the source on 21 March 2004. A period search was performed by \citet{zuritaheras2006}, and a pulsation of $414.8\pm0.5$ s was discovered. An orbital period of $9.737\pm0.004$ days was also reported by using the Rossi X-ray Timing Explorer (RXTE) proportional counter array data \citep{markwardt2003,corbet2005}. The source has a flat spectrum with $\Gamma=0.82^{+0.04}_{-0.04}$, which can also be fitted by a flat power law with an energy cutoff or a Comptonized model with $kT\sim5.5$ keV \citep{zuritaheras2006}. The spectrum shows a 6.4 keV iron line with an EW of $70^{+6}_{-7}$ eV. Previous studies indicate the source is a wind-fed accreting pulsar with a supergiant companion star. The source has been observed multiple times by \xmm, and we searched for pulsation in all the observations. The pulsations found in the different observations are consistent with each other within the 1$\sigma$ error values. The source is highly variable, and the flux of the source can vary from $2.19\times10^{-13}$ to $7.42\times10^{-11}$ erg s$^{-1}$ cm$^{-2}$. We noticed that whenever the source flux drops below $\sim5\times10^{-13}$ erg s$^{-1}$ cm$^{-2}$ the pulsation was undetectable.

The source 4XMM J174906.8--273233 was discovered in 1996 by ASCA. The source is also known as AXJ1749.1--2733 \citep{sakano2002}. In a 1995 ASCA observation, the source was not detected, and in 2003 INTEGRAL caught a short outburst, which indicates its transient nature \citep{grebenev2004}. \xmm first observed 4XMM J174906.8--273233 on 31 March 2007, and \citet{karasev2008} analyzed EPIC-pn data and detected a spin period of 132 s. The source was classified as a transient X-ray pulsar in a high-mass binary system. The source has been observed twice by \xmm in 2007 and 2008; however, the pulsation was only detected in the 2007 observation (ObsID: 0510010401). The non-detection of pulsations in 2008 could be due to the combination of two factors: (1) the source flux was almost an order of magnitude fainter than in the 2007 observation, and (2) the 2008 observation had a shorter exposure than the 2007 observation, which led to $\sim22$ times fewer net counts in the 2008 observation than in the 2007 observation.  The source spectrum is heavily absorbed and can be fitted by a steep power-law model with $\Gamma=1.3^{+0.1}_{-0.1}$; adding an iron line at 6.4 keV improves the fit minutely.

\subsection{Known IPs}
The source 4XMM J174517.0--321358 \citep{gong2022,vermette2023} was discovered by \chandra and serendipitously observed by \xmm in 2010. An iron-line emission complex and a pulsation of 614 s were detected using \xmm data. The source is classified as an IP with a WD of $0.8M_{\odot}$ \citep{vermette2023}. The source has been observed twice by \xmm, and in both observations, we detected a pulsation of 613 s. The X-ray spectrum looks like that of a typical IP with a flat spectral shape and iron emission complex.

The source 4XMM J174033.8--301501 was discovered by \suzaku in 2008 \citep{uchiyama2011b}. Later, the source was observed by \xmm on 18 March 2021 during a Galactic disk survey \citep{mondal2022}. The source spectrum is well described by emission from collisionally ionized diffuse gas with a plasma temperature of $\sim15.7$ keV plus an iron line emission complex. A period of 432.4 s was detected in both \suzaku and \xmm data. The source has been observed twice by \xmm in 2018 and 2021. In both \xmm observations, the detected pulsations are consistent. The source has a flat spectrum with $\Gamma=0.5^{+0.1}_{-0.1}$ and an Fe emission complex in the 6--7 keV band.

4XMM J174954.6--294336 was first discovered by \chandra \citep{jonker2014}. The source is classified as an IP based on the spin period of 1002 s and hard power-law spectral shape with complex iron line emission \citep{johnson2017,mondal2023}. This is only the second known IP that shows eclipses in X-rays. The source has been observed twice by \xmm, and the pulsation is not visible in ObsID 0801681401. \citet{mondal2023} discuss the possibility that the pulsation is suppressed due to a complex absorption behavior and the eclipse seen in the X-ray light curve.  

4XMM J174917.7--283329 is classified as IP \citep{mondal2023}. A period of 1212 s was detected in a 2017 \xmm observation. The continuum is best fitted by a partially absorbed apec model with a plasma temperature of $13$ keV. The source has been observed three times by \xmm, but the pulsation was detected only once, when the source flux was one order of magnitude higher than in the other two observations.

The source 4XMM J174816.9--280750 was observed by BeppoSAX during the GC survey in 1997--1998 \citep{sidoli2006}. The source has a spectrum with $\Gamma=1.3^{+0.6}_{-0.6}$ and strong emission lines at 6--7 keV, plus a coherent pulsation of period 593 s was found in \suzaku and \xmm data. These facts favor the source as an IP \citep{nobukawa2009}. The source has been observed ten times by \xmm, displaying significant variation in the pulsation period between different observations. A detailed, in-depth study of the source is required to determine whether the pulsation period variation is due to accretion or some other effects, such as the propeller phenomenon.

4XMM J174016.0--290337 was observed by \xmm on 29 September 2005 \citep{farrell2010}. The source displays Fe $K_{\alpha}$ emission and a periodic modulation with a period of 626 s. The source has been observed three times by \xmm, and in all cases, a pulsation period of 622 s is detected.

The source 4XMM J174009.0--28472 was first discovered by ASCA \citep{sakano2000} and a period of 729 s was found. The source was classified as NS pulsar based on the flat power-law-type spectrum shape \citep{sakano2000}. However, later near-infrared/optical studies suggested it is an IP \citep{kaur2010}. The source has been observed four times by \xmm, and we detected a similar pulsation period value in all observations. The source has a very flat spectrum $\Gamma=0.1^{+0.1}_{-0.1}$ with strong emission lines.

4XMM J174622.7--285218 is classified as an IP \citep{nucita2022}. The source was first observed in a \chandra observation of the GC, and a periodic signal of 1745 s was found \citep{muno2009}. The spectrum is characterized by $\Gamma=0.7^{+0.2}_{-0.2}$, and the 6.9 keV line is the strongest with an EW of $242^{+81}_{-74}$.

\subsection{Known polars}
The source 4XMM J174728.9--321441 was first observed by \chandra during the Galactic bulge survey \citep{jonker2014}. The source is classified as a polar based on its long period of 4860 s detected in X-rays and in \HeII $\lambda$5412 line emission \citep{wevers2017}. This source has the steepest spectrum in our sample with $\Gamma=1.8^{+0.4}_{-0.4}$, and no iron emission complex was detected in the \xmm spectrum. The non-detection of iron lines could be due to low signal-to-noise in the data.

\subsection{Unclassified sources}
Below, we try to classify the unknown sources using the scheme defined in Sect. \ref{sec:catagory_class}. This is a tentative classification; further follow-up of the individual sources is required to constrain their true nature. For many sources, we do not have any \gaia counterpart; therefore, the estimation of the distance to the source using parallax was not possible and hence the luminosity is not calculated. In such a case, we only used the first three criteria for classification.
\subsubsection{Likely NS HMXRBs}
The only two sources matching the NS HMXRB criteria are XMMU J175441.9--265919 and 4XMM J175525.0--260402. Both have relatively high HR values: $0.02\pm0.06$ and $0.05\pm0.02$, respectively. The upper limits on the EWs of the 6.7 keV line for J175441.9 and J175525.0 are $\sim28$ eV and $\sim34$ eV, respectively. The spin periods of these two systems are 1.36 s and 392.5 s. The source J175525.0 was detected three times by \xmm; however, the pulsation was detected only in the longest observation. A luminosity estimation was not possible as we did not find any counterparts in \gaia catalogs.
\subsubsection{Likely IPs}
The sources we categorize as IP are 4XMM J173058.9--350812, 4XMM J175301.3--291324, 4XMM J175740.5--285105, and 4XMM J175511.6--260315. The periods found from these systems are below 2500 s and the HRs are above -0.2. The 6.7 keV line EWs for the sources J173058.9, J175301.3, J175740.5, and J175511.6 are $236^{+105}_{-83}$, $105^{+189}$, $112^{+61}_{-41}$, and $194^{+162}_{-188}$ eV, respectively. The source J175301.3 was observed three times by \xmm, and a $\sim672$ s period was consistently found in two of those observations. The source J175511.6 was detected three times by \xmm; however, the period of $\sim1135$ s was detected only in the longest observation. We detected a likely \gaia counterpart for the sources J173058.9 and J175511.6 and the distances estimated from their parallaxes are $3.2^{+2.2}_{-1.3}$ and $4.9^{+3.2}_{-2.4}$ kpc, respectively. The luminosity of these two sources is in the range $(0.8-1.9)\times10^{33}$ erg s$^{-1}$, which is typical for accreting magnetic CVs \citep{suleimanov2022}.
\subsubsection{Likely polars}
The sources that are likely to be polars are 4XMM J173837.0--304818, 4XMM J175327.8--295716, 4XMM J175244.4--285851, 4XMM J175328.4--244627, and XMMU J180140.3--234221. These sources have very low HR values and long periods (see Fig. \ref{fig:period_hr}), which suggests that these sources are most likely to be polars. The long periods are most likely associated with the synchronized spin-orbital period of the WDs. All these sources have relatively soft spectra of photon index $\Gamma=0.8-2.1$. For most of the polar-type sources, we have an upper limit on the EW of the 6.7 keV emission line. This is primarily because these sources have very low net counts that give
$\leq50$ bins in the 0.2–10 keV spectrum. The source J175328.4 is the brightest in the polar sample and has the best signal-to-noise spectrum compared to the other four sources. In this case, the EW of the 6.7 keV emission line is $28^{+26}_{-28}$ eV, which is much smaller than the typical values found in IPs. The source J175327.8 was observed six times by \xmm; however, the periodicity was significantly detected only in the two observations that have an exposure above 25 ks. We found a likely \gaia counterpart for the source J175328.4 and its estimated distance is $1.8^{+0.4}_{-0.2}$ kpc, giving a luminosity of $2\times10^{33}$ erg s$^{-1}$.
\subsubsection{Unknowns}
We classify the sources  XMMU J173029.8--330920, 4XMM J174809.8--300616, and 4XMM J175452.0--295758 as unknowns. These sources have high HR and periods similar to IPs. However, the 6.7 keV line was not detected clearly and we could only set an upper limit on its EW. The EW of the 6.7 keV line for the sources J173029.8, J174809.8, and J175452.0 are <160, <172, and <206 eV, respectively. The source J174809.8 has been observed twice by \xmm. However, it is relatively faint, with a flux of a few times $10^{-13}$ erg cm$^{-2}$ s$^{-1}$ and therefore the period was detected only in the longer observation. The source J175452.0 was detected three times by \xmm; however, the pulsation was detected only in the longest observation.

\subsubsection{NS or WD?}
The compact object in 4XMM J174445.6--271344 is not clearly identified. Also known as HD 161103, it was observed by \xmm on 26 January 2004. \citet{LopesdeOliveira2006} did a detailed multiwavelength spectroscopic study of this source and suggested that the system hosts an NS; however, a WD scenario was not excluded. From optical spectroscopy, the companion star of this system is recognized as a Be star. We detected a periodicity of 3196 s from the X-ray light curve. The X-ray spectra show strong 6.4, 6.7, and 6.9 keV emission lines with EWs of $80^{+39}_{-40}$, $371^{+88}_{-76}$, and $109^{+43}_{-47}$ eV, respectively. Such strong 6.7 and 6.9 keV emission lines are not typically seen in accreting NS HMXRBs. Also, the source has a much softer spectrum ($\rm HR=-0.42\pm0.03$ in Fig. \ref{fig:period_hr}) than the two confirmed NS HMXRBs (4XMM J172511.3--361657 and 4XMM J172511.3--361657) in our sample.

\section{Conclusion}
We systematically searched for periodic X-ray sources in the inner Galactic disk, which extends from $l\sim350\degr$ to $l\sim+7\degr$ and includes the GC, using \xmm Heritage observations and archival data. We find 26 sources that show periodicity in their X-ray light curves, of which 12 have previously reported periods.   For these 12 sources, we have obtained periods consistent with those previously reported. We have detected the periodicity in the other 14 sources for the first time. We classified the sources based on the values of the HR, period, and iron emission complex in the 6--7 keV band. Of these 14 sources, we classify two as NS X-ray binaries, four as likely IPs, five as likely polars, and three as unknowns. The IP-type sources display a steep X-ray spectrum with $\Gamma\leq1.1$ and an iron emission complex in the 6--7 keV band. The spectra of polars are much softer compared to IPs. 

\begin{table}
\caption{Sources with possible \gaia optical counterparts.}
\label{table:list_tab_optical}
\setlength{\tabcolsep}{2.0pt}                   
\renewcommand{\arraystretch}{1.5}               
\centering
\begin{tabular}{c c c c c c}
\hline\hline
\multirow{2}{*}{XMM Name} & Density & \multirow{2}{*}{$G_{\rm mag}$} & Plx & Distance & $L_{\rm x}$\\
& arcsec$^{-2}$ && mas & kpc & erg s$^{-1}$\\
\hline

4XMM J173058.9 & 0.009 & 20.05 & $0.59$ & $3.2^{+2.2}_{-1.3}$ & $1.9\times10^{33}$\\ \hline
4XMM J174033.8 & 0.009 & 19.23 & $0.23$ & $4.3^{+1.5}_{-1.3}$ & $6.4\times10^{33}$\\ \hline
4XMM J174009.0 & 0.03 & 18.79 & $0.71$ & $1.8^{+1.2}_{-0.5}$ & $2.3\times10^{33}$\\ \hline
4XMM J174954.6 & 0.1 & 18.97 & $0.61$ & $1.6^{+2.8}_{-2.5}$ & $5.3\times10^{32}$\\ \hline
4XMM J174816.9 & 0.01 & 21.13 & $0.63$ & $2.9^{+2.3}_{-1.4}$ & $8.6\times10^{32}$\\ \hline 
4XMM J175511.6 & 0.02 & 18.64 & $0.40$ & $4.9^{+3.2}_{-2.4}$ & $8.8\times10^{32}$\\ \hline
4XMM J175328.4 & 0.02 & 8.40 & $0.55$ & $1.8^{+0.4}_{-0.2}$ & $2.2\times10^{33}$\\ \hline

\end{tabular}
\tablefoot{Sources with \gaia counterparts found within the 3$\sigma$ positional error of \xmm. The density was calculated by drawing a circle with a radius of 1 arcmin at the source position. }
\end{table}

\begin{acknowledgements}
SM and GP acknowledge financial support from the European Research Council (ERC) under the European Union’s Horizon 2020 research and innovation program HotMilk (grant agreement No. 865637). SM and GP also acknowledge support from Bando per il Finanziamento della Ricerca Fondamentale 2022 dell’Istituto Nazionale di Astrofisica (INAF): GO Large program and from the Framework per l’Attrazione e il Rafforzamento delle Eccellenze (FARE) per la ricerca in Italia (R20L5S39T9). KM is partially supported by the NASA ADAP program (NNH22ZDA001N-ADAP). We thank the referee for the comments, corrections, and suggestions that significantly improved the manuscript.

\end{acknowledgements}

%
%

\bibliographystyle{aa} 
\bibliography{refs}

\begin{appendix}

\section{Additional figures and tables}


\begin{landscape}
\begin{table}
\caption{Various details of the X-ray pulsators in the GC plus Galactic disk.}
\label{table:list_tab}
\setlength{\tabcolsep}{2.5pt}                   
\renewcommand{\arraystretch}{1.5}               
\centering
\begin{tabular}{c c c c c c c c c c c c}
\hline\hline
\multirow{2}{*}{lat} & \multirow{2}{*}{lon} & \multirow{2}{*}{ObsID} & $\rm Pos_{err}$ & \multicolumn{2}{c}{Period (s)} & PF & \multirow{2}{*}{XMM Name} & \multirow{2}{*}{P$_\gaia$} & Sig & \multirow{2}{*}{Type} & \multirow{2}{*}{References}\\ 
&&& arcsec & Our & Previous & 2--10 keV & & & >$\sigma$\\ \hline

351.4972 & -0.3537 & 0886070601 & 0.14 & $414.5\pm1.4$ & 414.8 & $68.4\pm1.6$ & 4XMM J172511.3--361657 & $4\times10^{-4}$ & 3 & NS HMXRB & \citet{zuritaheras2006}\\ \hline 
353.1032 & -0.6956 & 0861171201 & 0.73 & $607.5\pm3.7$ && $45.6\pm10.8$ & 4XMM J173058.9--350812 & $1.5\times10^{-2}$ & 3 & Likely IP & \\ \hline
354.7018 & 0.4782 & 0916800201 & 0.35 & $517.6\pm3.6$ && $41.3\pm12.5$ & XMMU J173029.8--330920 & $1\times10^{-3}$ & 3 & Unknown & Mondal et al. in prep\\ \hline
357.1486 & -1.6563 & 0865510101 & 0.48 & $614.2\pm1.4$ & 614 & $28.2\pm4.5$ & 4XMM J174517.0--321358 & $4\times10^{-3}$ & 3 & IP & \citet{vermette2023}\\ \hline 
357.3792 & -2.0600 & 0743980401 & 0.57 & $4768.3\pm151.5$ & 4860 & $74.2\pm22.4$ & 4XMM J174728.9--321441 & $2.1\times10^{-2}$ & 3 & Polar & \citet{wevers2017}\\ \hline
357.6116 & 0.3037 & 0886020101 & 0.79 & $5067.3\pm260.5$ && $57.9\pm16.2$ & 4XMM J173837.0--304818 & $1.2\times10^{-2}$ & 3 & Likely Polar & \\ \hline
358.3043 & 0.2442 & 0886010601 & 0.48 & $432.4\pm1.9$ & 432.1 & $54.8\pm7.7$ & 4XMM J174033.8--301501 & $6.8\times10^{-3}$ & 3 & IP & \citet{mondal2022}\\ \hline 
359.2786 & 0.9300 & 0764191201 & 0.17 &  $623.2\pm2.6$ & 626 & $37.2\pm3.8$ & 4XMM J174016.0--290337 & $1.4\times10^{-3}$ & 3 & IP & \citet{farrell2010}\\ \hline
359.2882 & -1.0793 & 0152920101 & 0.65 & $2179.6\pm18.2$ && $64.1\pm19$ & 4XMM J174809.8--300616 & $2.5\times10^{-2}$ & 3 & Unknown & \\ \hline
359.4941 & 1.0946 & 0764191101 & 0.6 & $725.9\pm3.5$ & 729 & $45.1\pm6.8$ & 4XMM J174009.0--284725 & $3.2\times10^{-2}$ & 3 & IP & \citet{kaur2010}\\ \hline 
359.8061 & -1.2096 & 0801683401 & 0.56 & $997.7\pm7.6$ & 1001.5 & $31\pm8.9$ & 4XMM J174954.6--294336 & $7.7\times10^{-2}$ & 3 & IP & \citet{mondal2023}\\ \hline
0.0036 & -1.9883 & 0801682901 & 0.76 & $2917.0\pm68.5$ && $54.9\pm12.7$ & 4XMM J175327.8--295716 & $1.8\times10^{-1}$ & 2 & Likely Polar & \\ \hline 
0.1413 & -0.1089 & 0762250301 & 0.32 & $1737.7\pm5.4$ & 1745 & $15.8\pm3.7$ & 4XMM J174622.7--285218 & $2.5\times10^{-3}$ & 2 & IP & \citet{muno2009}\\ \hline 
0.1476 & -2.2564 & 0402280101 & 0.59 & $552.1\pm1.4$ && $45.1\pm9.8$ & 4XMM J175452.0--295758 & $9.9\times10^{-2}$ & 3 & Unknown & \\ \hline
0.5849 & -1.5353 & 0801682801 & 0.49 & $672.5\pm3.0$ && $77.8\pm18.9$ & 4XMM J175301.3--291324 & $7.7\times10^{-2}$ & 3 & Likely IP & \\ \hline
0.7407 & -0.4932 & 0801681301 & 0.44 & $1209.7\pm11.7$ & 1212.4 & $44.3\pm8.3$ & 4XMM J174917.7--283329 & $1.3\times10^{-2}$ & 3 & IP & \citet{mondal2023}\\ \hline
0.7632 & -1.3587 & 0801682601 & 0.97 & $6784.4\pm439.5$ && $73\pm20.4$ & 4XMM J175244.4--285851 & $2.8\times10^{-1}$ & 3 & Likely Polar & \citet{bahramian2021}\\ \hline 
0.9918 & -0.0821 & 0783160101 & 0.47 & $593.6\pm0.7$ & 593 & $41.3\pm7.3$ & 4XMM J174816.9--280750 & $6.8\times10^{-2}$ & 3 & IP & \citet{sidoli2006}\\ \hline 
1.3573 & 1.0522 & 0201200101 & 1.9 & $3195.8\pm116.3$ & 3200 & $28.9\pm6.5$ & 4XMM J174445.6--271344 & $6.2\times10^{-1}$ & 3 & NS/WD ? & \citet{LopesdeOliveira2006}\\ \hline 
1.4196 & -2.2266 & 0782770201 & 0.51 & $354.68\pm0.7$ && $47.6\pm10.9$ & 4XMM J175740.5--285105 & $9.2\times10^{-1}$ & 3 & Likely IP & \\ \hline
1.5909 & 0.0633 & 0510010401 & 1.9 & $132.1\pm0.3$ & 132 & $28.9\pm3.4$ & 4XMM J174906.8--273233 & $2.0\times10^{-1}$ & 3 & NS HMXRB & \citet{karasev2008}\\ \hline
2.6999 & -0.7212 & 0886081101 & 0.59 & $1.36652\pm2\times10^{-5}$ && $48.1\pm12.6$ & XMMU J175441.9--265919 & $7.7\times10^{-3}$ & 3 & Likely NS XRB & \\ \hline
3.5626 & -0.3453 & 0886081301 & 0.53 & $1134.5\pm12.4$ && $79.1\pm14.4$ & 4XMM J175511.6--260315 & $1.7\times10^{-2}$ & 3 & Likely IP & \\ \hline
3.5766 & -0.3953 & 0886081301 & 0.51 & $392.5\pm1.5$ && $28.2\pm5$ & 4XMM J175525.0--260402 & $1.5\times10^{-2}$ & 3 & Likely NS XRB & \\ \hline
4.4701 & 0.6376 & 0840910501 & 0.36 & $4093.3\pm182.3$ && $34.8\pm6.1$ & 4XMM J175328.4--244627 & $8.8\times10^{-2}$ & 3 & Likely Polar & \\ \hline
6.3308 & -0.4448 & 0886110501 & 0.64 & $5215.9\pm7.5$ && $61.8\pm19.3$ & XMMU J180140.3--234221 & $1.7\times10^{-2}$ & 2 & Likely Polar & \\ \hline

\end{tabular}
\tablefoot{The details of the X-ray pulsators, including the positional information, \xmm detection ID, and the source association in the 4XMM catalog. The source XMMU J173029.8--330920, XMMU J175441.9--265919, and XMMU J180140.3--234221 are the first time detected by \xmm in our ongoing \emph{Heritage} survey of the Galactic disk. We also searched for \swift-XRT and \chandra counterparts of these two sources, but no counterparts were found. The P$_\gaia$ represents the probability of spurious association with a \gaia source.}
\end{table}
\end{landscape}

\begin{table*}
\caption{Details of the pulsators for which more than one observation is available.}
\label{table:list_tab1}
\setlength{\tabcolsep}{4pt}                   
\renewcommand{\arraystretch}{1.47}               
\centering
\begin{tabular}{c c c c c c}
\hline\hline
\multirow{2}{*}{XMM Name} & \multirow{2}{*}{ObsID} & \multirow{2}{*}{Date} & Flux & \multirow{2}{*}{Period} & Exposure\\ 
&&& erg s$^{-1}$ cm$^{-2}$ && ks\\
\hline

\multirow{10}{*}{4XMM J172511.3--361657} 
& 0206380401 & 2004-03-21 & $(7.42\pm0.04)\times10^{-11}$ & $413.8\pm3.7$ & 10.9\\
& 0405640201 & 2006-08-29 & $(3.94\pm0.17)\times10^{-13}$ & & 22.9\\
& 0405640301 & 2006-08-31 & $(6.46\pm0.04)\times10^{-11}$ & $414.5\pm3.8$ & 11.3\\
& 0405640401 & 2006-09-04 & $(2.55\pm0.02)\times10^{-11}$ & $414.4\pm3.3$ & 12.5\\
& 0405640501 & 2006-09-06 & $(3.21\pm0.08)\times10^{-12}$ & $409.8\pm3.5$ & 11.9\\
& 0405640601 & 2006-09-08 & $(5.53\pm0.27)\times10^{-13}$ & & 13.9\\
& 0405640701 & 2006-09-15 & $(1.81\pm0.02)\times10^{-11}$ & $414.4\pm1.7$ & 22.9\\
& 0405641001 & 2006-09-27 & $(2.19\pm0.21)\times10^{-13}$ & & 12.4\\
& 0405640901 & 2006-09-28 & $(2.77\pm0.02)\times10^{-11}$ & $413.9\pm2.6$ & 15.2\\ 
& 0405640801 & 2006-10-01 & $(4.12\pm0.02)\times10^{-11}$ & $413.2\pm2.5$ & 15.7\\
& 0886070601 & 2006-10-01 & $(3.56\pm0.03)\times10^{-11}$ & $414.5\pm1.4$ & 26.6 \\
\hline

\multirow{3}{*}{4XMM J174517.0--321358} 
& 0553950201 & 2010-10-09 & $(2.10\pm0.06)\times10^{-12}$ & $613.8\pm0.9$ & 86.4\\
& 0870990201 & 2021-02-28 & $(1.24\pm0.02)\times10^{-12}$ & $614.2\pm2.4$ & 31.6\\
& 0865510101 & 2021-03-02 & $(1.26\pm0.02)\times10^{-12}$ & $614.2\pm1.4$ & 62.9\\ 
\hline

\multirow{2}{*}{4XMM J174033.8--301501} 
& 0823030101 & 2018-09-29 & $(1.96\pm0.09)\times10^{-12}$ & $433.0\pm5.6$ & 8.0\\
& 0886010601 & 2021-03-18 & $(2.88\pm0.06)\times10^{-12}$ & $432.4\pm1.9$ & 23.0\\ 
\hline

\multirow{3}{*}{4XMM J174016.0--290337} 
& 0304220101 & 2005-09-29 & $(3.61\pm0.13)\times10^{-12}$ & $624.8\pm9.4$ & 8.5\\
& 0764191201 & 2016-03-05 & $(5.14\pm0.06)\times10^{-12}$ & $623.2\pm2.6$ & 33\\
& 0764191101 & 2016-03-05 & $(8.78\pm0.30)\times10^{-12}$ & $622.6\pm2.6$ & 33\\ 
\hline

\multirow{2}{*}{4XMM J174809.8--300616} 
& 0152920101 & 2003-04-02 & $(2.96\pm0.17)\times10^{-13}$ & $2179.6\pm18.2$ & 52.2\\
& 0801683301 & 2018-04-06 & $(2.08\pm0.39)\times10^{-13}$ && 29.8 \\ 
\hline

\multirow{4}{*}{4XMM J174009.0--284725} 
& 0511010701 & 2008-02-27 & $(3.75\pm0.09)\times10^{-12}$ & $733.1\pm14.6$ & 9.3\\
& 0764191501 & 2016-02-25 & $(5.89\pm0.22)\times10^{-12}$ & $725.5\pm3.7$ & 30.5\\ 
& 0764191101 & 2016-03-05 & $(5.90\pm0.14)\times10^{-12}$ & $725.9\pm3.5$ & 33\\
& 0764191601 & 2016-03-10 & $(5.63\pm0.21)\times10^{-12}$ & $725.5\pm6.0$ & 19\\
\hline

\multirow{2}{*}{4XMM J174954.6--294336} 
& 0801681401 & 2017-10-07 & $(1.28\pm0.05)\times10^{-12}$ & & 28\\
& 0801683401 & 2018-04-06 & $(1.73\pm0.06)\times10^{-12}$ & $997.7\pm7.6$ & 29.2\\
\hline

\multirow{6}{*}{4XMM J175327.8–295716}
& 0085580501 & 2000-10-11 & $(3.35\pm0.48)\times10^{-13}$ && 8.0\\
& 0085581501 & 2001-03-24 & $(1.93\pm1.22)\times10^{-13}$ && 7.5\\
& 0085581601 & 2001-09-07 & $(3.77\pm1.07)\times10^{-13}$ && 8.2\\
& 0085581801 & 2002-03-13 & $(1.21\pm0.53)\times10^{-13}$ && 8.2\\
& 0801682901 & 2018-09-07 & $(4.29\pm0.38)\times10^{-13}$ & $2917.0\pm68.5$ & 27.9\\
& 0801683601 & 2018-09-25 & $(3.05\pm0.31)\times10^{-13}$ & $2870.2\pm58.1$ & 31.7\\ 
\hline

\multirow{3}{*}{4XMM J175452.0--295758}
& 0085580501 & 2000-10-11 & $(1.88\pm0.45)\times10^{-13}$ && 8.0\\
& 0206590201 & 2004-09-05 & $(2.46\pm0.27)\times10^{-13}$ && 20.9\\
& 0402280101 & 2006-09-10 & $(2.68\pm0.19)\times10^{-13}$ & $552.1\pm1.4$ & 44.1\\
\hline

\multirow{3}{*}{4XMM J175301.3--291324}
& 0801682501 & 2018-09-03 & $(3.29\pm1.29)\times10^{-13}$ && 29.0\\
& 0801682801 & 2018-09-09 & $(1.86\pm0.11)\times10^{-13}$ & $673.1\pm3.1$ & 32.9\\
& 0801683501 & 2018-09-25 & $(4.12\pm0.57)\times10^{-13}$ & $672.5\pm3.0$ & 31.5\\
\hline

\end{tabular}
\tablefoot{The \xmm ObsIDs details for sources in which more than one observation is available. The flux is taken from the 4XMM catalog. In many cases, the pulsation was not detected if the exposure was short or the source flux was below a certain limit.}
\end{table*}

\begin{table*}
\caption{Table \ref{table:list_tab1} Continued.}
\label{table:list_tab2}
\setlength{\tabcolsep}{4pt}                   
\renewcommand{\arraystretch}{1.5}               
\centering
\begin{tabular}{c c c c c c}
\hline\hline
XMM Name & ObsID & Date & Flux & Period & Exposure\\ \hline

\multirow{3}{*}{4XMM J174917.7--283329}
& 0410580401 & 2006-09-22 & $(1.85\pm0.50)\times10^{-13}$ && 32.9\\
& 0410580501 & 2006-09-26 & $(3.07\pm0.63)\times10^{-13}$ && 32.4\\
& 0801681301 & 2017-10-07 & $(1.32\pm0.03)\times10^{-12}$ & $1209.7\pm11.7$ & 28.0 \\
\hline

\multirow{10}{*}{4XMM J174816.9--280750}
& 0112970101 & 2000-09-23 & $(7.89\pm0.43)\times10^{-13}$ & $595.1\pm5.5$ & 16.3 \\
& 0112970201 & 2000-09-23 & $(9.34\pm0.79)\times10^{-13}$ & $587.8\pm4.5$ & 18.1\\ 
& 0144220101 & 2003-03-12 & $(9.42\pm0.61)\times10^{-13}$ & $592.9\pm1.4$ & 52.4 \\
& 0205240101 & 2005-02-26 & $(6.37\pm0.19)\times10^{-13}$ & $592.7\pm1.4$ & 51\\
& 0694640801 & 2012-10-06 & $(8.04\pm0.37)\times10^{-13}$ & $590.9\pm1.7$ & 41.9\\
& 0694641501 & 2012-10-06 & $(6.67\pm0.17)\times10^{-13}$ & $592.9\pm1.4$ & 51.8\\
& 0694640701 & 2012-10-02 & $(6.47\pm0.19)\times10^{-13}$ & $592.5\pm1.7$ & 44.4\\
& 0694641401 & 2012-09-30 & $(7.43\pm0.41)\times10^{-13}$ && 51.8\\
& 0783160101 & 2016-10-02 & $(8.51\pm0.23)\times10^{-13}$ & $593.6\pm0.7$ & 106\\
& 0862471201 & 2020-10-04 & $(1.01\pm0.05)\times10^{-12}$ & & 46.9\\
\hline

\multirow{2}{*}{4XMM J174445.6--271344}
& 0201200101 & 2004-02-26 & $(2.03\pm0.04)\times10^{-12}$ & $3195.8\pm116.3$ & 17.8\\
& 0691760101 & 2012-09-08 & $(1.33\pm0.02)\times10^{-12}$ && 22.9\\
\hline

\multirow{2}{*}{4XMM J174906.8--273233}
& 0510010401 & 2007-03-31 & $(1.13\pm0.01)\times10^{-11}$ & $132.1\pm0.3$ & 12.2 \\
& 0511010301 & 2008-03-04 & $(2.44\pm0.12)\times10^{-12}$ & & 8.9\\
\hline

\multirow{3}{*}{4XMM J175511.6--260315}
& 0148090101 & 2003-03-17 & $(1.07\pm0.14)\times10^{-12}$ & & 12.1\\
& 0148090501 & 2003-09-11 & $(1.53\pm0.17)\times10^{-12}$ & & 11.2\\
& 0886081301 & 2023-04-06 & $(3.06\pm0.21)\times10^{-12}$ & $1134.5\pm12.4$ & 24\\
\hline

\multirow{3}{*}{4XMM J175525.0--260402}
& 0148090101 & 2003-03-17 & $(3.03\pm0.30)\times10^{-12}$ & & 12.1\\
& 0148090501 & 2003-09-11 & $(4.47\pm0.43)\times10^{-12}$ & & 11.2\\
& 0886081301 & 2023-04-06 & $(1.84\pm0.10)\times10^{-12}$ & $392.5\pm1.5$ & 24\\
\hline

\end{tabular}
\tablefoot{Same columns as Table \ref{table:list_tab1}.}
\end{table*}

\begin{landscape}
\begin{table}
\caption{Details of the spectral fit.}
\label{table:spec_tab}
\setlength{\tabcolsep}{5.0pt}                   
\renewcommand{\arraystretch}{1.5}               
\centering
\begin{tabular}{c c c c c c c c c c c c}
\hline\hline
\multirow{2}{*}{XMM Name} & $N_{\rm H}$ & \multirow{2}{*}{$\Gamma$} & \multirow{2}{*}{$N_{\rm po}$} & \multirow{2}{*}{$N_{\rm 6.4}$} & $\rm EW_{6.4}$ & \multirow{2}{*}{$N_{\rm 6.7}$} & $\rm EW_{6.7}$ & \multirow{2}{*}{$N_{\rm 6.9}$} & $\rm EW_{6.9}$ & \multirow{2}{*}{$\chi^2/\rm d.o.f$} & Flux\\ 
& $\times10^{22}\rm\ cm^{-2}$ &&&& eV && eV && eV && 0.2--10 keV\\ \hline
4XMM J172511.3 & $10.2^{+0.3}_{-0.3}$ & $0.82^{+0.04}_{-0.04}$ & $3.1^{+0.3}_{-0.3}\times10^{-3}$ & $3.9^{+0.6}_{-0.6}\times10^{-5}$ & $70^{+6}_{-7}$ & $<5\times10^{-6}$ & $<8$ & $<3\times10^{-6}$ & $<7$ & 2364/2299 & $3.56\times10^{-11}$\\ \hline
4XMM J173058.9 & $1.4^{+0.6}_{-0.4}$ & $0.6^{+0.3}_{-0.2}$ & $5^{+3}_{-2}\times10^{-5}$ & $3^{+3}_{-2}\times10^{-6}$ & $129^{+70}_{-58}$ & $5^{+3}_{-3}\times10^{-6}$ & $236^{+105}_{-83}$ & $<1\times10^{-6}$ & $<50$ & 69/82 & $1.52\times10^{-12}$\\ \hline
XMMU J173029.8 & $5^{+5}_{-4}$ & $0.0^{+0.7}_{-0.6}$ & $9^{+28}_{-6}\times10^{-6}$ & $2^{+1}_{-1}\times10^{-6}$ & $270^{+135}_{-135}$ & $<2\times10^{-6}$ & $<160$ & $2^{+3}_{-1}\times10^{-6}$ & $204^{+306}_{-102}$ & 11/17 & $6.70\times10^{-13}$\\ \hline
4XMM J174517.0 & $2.8^{+0.3}_{-0.3}$ & $1.1^{+0.1}_{-0.1}$ & $9^{+2}_{-2}\times10^{-5}$ & $2.8^{+0.8}_{-0.8}\times10^{-6}$ & $192_{-31}^{+32}$& $1.4^{+0.8}_{-0.8}\times10^{-6}$ & $78^{+34}_{-19}$ & $1.7^{+0.8}_{-0.8}\times10^{-6}$ & $129^{+51}_{-35}$ & 339/326 & $9.59\times10^{-13}$\\ \hline
4XMM J174728.9 & $0.2^{+0.1}_{-0.1}$ & $1.8^{+0.4}_{-0.4}$ & $7^{+4}_{-2}\times10^{-6}$ & $<4\times10^{-7}$ & $<1500$ & $<5\times10^{-7}$ & $<2000$ & $<5\times10^{-7}$ & $<2200$ & 16/28 & $4.00\times10^{-14}$\\ \hline
4XMM J173837.0 & $0.9^{+0.7}_{-0.5}$ & $0.8^{+0.4}_{-0.4}$ & $1.1^{+1.1}_{-0.5}\times10^{-5}$ & $<8\times10^{-7}$ & $<351$ & $7^{+91}_{-7}\times10^{-8}$ & $29^{+329}_{-29}$ & $1^{+9}_{-1}\times10^{-7}$ & $59^{+320}_{-59}$ & 23/42 & $2.06\times10^{-13}$\\ \hline
4XMM J174033.8 & $0.9^{+0.2}_{-0.2}$ & $0.5^{+0.1}_{-0.1}$ & $6^{+1}_{-1}\times10^{-5}$ & $7^{+2}_{-2}\times10^{-6}$ & $226^{+50}_{-46}$ & $4^{+2}_{-2}\times10^{-6}$ & $103^{+42}_{-40}$ & $3^{+2}_{-2}\times10^{-6}$ & $95^{+50}_{-49}$ & 185/189 & $2.88\times10^{-12}$\\ \hline
4XMM J174016.0 & $0.44^{+0.03}_{-0.03}$ & $0.63^{+0.04}_{-0.02}$ & $1.36^{+0.09}_{-0.07}\times10^{-4}$ & $9^{+2}_{-2}\times10^{-6}$ & $182^{+34}_{-20}$ & $6^{+2}_{-2}\times10^{-6}$ & $91^{+23}_{-16}$ & $7^{+2}_{-2}\times10^{-6}$ & $135^{+27}_{-29}$ & 736/607 & $5.14\times10^{-12}$\\ \hline
4XMM J174809.8 & $2^{+2}_{-1}$ & $0.3^{+0.5}_{-0.5}$ & $6^{+9}_{-3}\times10^{-6}$ & $5^{+7}_{}\times10^{-7}$ & $175^{+71}$ & $<7\times10^{-7}$ & $<172$ & $3^{+14}_{-3}\times10^{-7}$ & $99^{+264}_{-99}$ & 50/48 & $2.96\times10^{-13}$\\ \hline
4XMM J174009.0 & $0.8^{+0.3}_{-0.2}$ & $0.1^{+0.1}_{-0.1}$ & $6^{+2}_{-1}\times10^{-5}$ & $1.8^{+0.5}_{-0.5}\times10^{-5}$ & $359^{+65}_{-120}$ & $1.6^{+0.6}_{-0.6}\times10^{-5}$ & $265^{+78}_{-115}$ & $7^{+6}_{-6}\times01^{-6}$ & $109^{+66}_{-56}$ & 170/166 & $5.90\times10^{-12}$\\ \hline
4XMM J174954.6 & $2.4^{+0.9}_{-0.7}$ & $0.4^{+0.3}_{-0.3}$ & $3^{+2}_{-1}\times10^{-5}$ & $2^{+1}_{-1}\times10^{-6}$ & $107^{+69}_{-59}$& $3^{+2}_{-2}\times10^{-6}$ & $172^{+103}_{-85}$ & $<3^{}_{}\times10^{-6}$ & $<150$ & 90/91 & $1.73\times10^{-12}$\\ \hline
4XMM J175327.8 & $7^{+7}_{-5}\times10^{-2}$ & $1.3^{+0.3}_{-0.3}$ & $2.8^{+0.9}_{-0.7}\times10^{-5}$ & $<2\times10^{-6}$ & $<774$ & $<2\times10^{-6}$ & $<660$ & $<2\times10^{-6}$ & $<751$ & 64/41 & $4.29\times10^{-13}$\\ \hline
4XMM J174622.7 & $4.3^{+0.9}_{-0.7}$ & $0.7^{+0.2}_{-0.2}$ & $2.0^{+1.0}_{-0.6}\times10^{-5}$ & $1^{+5}_{-1}\times10^{-7}$ & $20^{+46}_{-20}$ & $1.1^{+0.6}_{-0.6}\times10^{-7}$ & $207^{+76}_{-71}$ & $2^{+6}_{-2}\times10^{-7}$ & $30^{+67}_{-30}$ & 192/187 & $5.13\times10^{-13}$\\ \hline
4XMM J175452.0 & $2^{+3}_{-1}$ & $0.9^{+0.9}_{-0.7}$ & $1.2^{+4.4}_{-0.9}\times10^{-5}$ & $1.4^{+0.7}_{-0.7}\times10^{-6}$ & $690^{+445}_{-445}$ & $<9\times10^{-7}$ & $<206$ & $<1\times10^{-6}$ & $<529$ & 33/41 & $2.68\times10^{-13}$\\ \hline
4XMM J175301.3 & $0.3^{+0.5}_{-0.2}$ & $0.5^{+0.3}_{-0.3}$ & $5^{+3}_{-2}\times10^{-6}$ & $10^{+6}_{-6}\times10^{-7}$ & $521^{+288}_{-225}$ & $4^{+8}_{-4}\times10^{-7}$ & $105^{+189}_{-105}$ & $4^{+10}_{-4}\times10^{-7}$ & $179^{+387}_{-179}$ & 63/54 & $1.86\times10^{-13}$\\ \hline
4XMM J174917.7 & $3.2^{+0.7}_{-0.6}$ & $1.0^{+0.2}_{-0.2}$ & $9^{+5}_{-3}\times10^{-5}$ & $2^{+1}_{-1}\times10^{-6}$ & $94^{+50}_{-47}$ & $4^{+2}_{-2}\times10^{-6}$ & $246^{+88}_{-67}$ & $9^{+16}_{-9}\times10^{-7}$ & $52^{+112}_{-52}$ & 107/108 & $1.31\times10^{-12}$\\ \hline
4XMM J175244.4 & $0.1^{+0.2}_{-0.1}$ & $0.7^{+0.3}_{-0.3}$ & $7^{+3}_{-2}\times10^{-6}$ & $<8\times10^{-7}$ & $<388$ & $6^{+95}_{-6}\times10^{-8}$ & $17^{+231}_{-17}$ & $1.4^{+0.9}_{-0.9}\times10^{-6}$ & $796^{+443}_{-405}$ & 29/51 & $2.11\times10^{-13}$\\ \hline
4XMM J174816.9 & $27^{+7}_{-6}$ & $1.3^{+0.6}_{-0.6}$ & $2.0^{+4}_{-1}\times10^{-4}$ & $1.8^{+0.7}_{-0.7}\times10^{-6}$ & $221^{+86}_{-86}$ & $1.0^{+0.7}_{-0.7}\times10^{-6}$ & $109^{+76}_{-76}$ & $6^{+9}_{-6}\times10^{-7}$ & $71^{+107}_{-71}$ & 159/164 & $8.50\times10^{-13}$\\ \hline
4XMM J174445.6 & $0.37^{+0.02}_{-0.02}$ & $1.75^{+0.06}_{-0.06}$ & $3.3^{+0.2}_{-0.2}\times10^{-4}$ & $2^{+1}_{-1}\times10^{-6}$ & $80^{+39}_{-40}$ & $7^{+2}_{-2}\times10^{-6}$ & $371^{+88}_{-76}$ & $2^{+1}_{-1}\times10^{-6}$ & $109^{+43}_{-47}$ & 399/412 & $2.03\times10^{-12}$\\ \hline
4XMM J175740.5 & $3^{+2}_{-1}$ & $0.2^{+0.4}_{-0.3}$ & $1.1^{+1.1}_{-0.5}\times10^{-5}$ & $1.6^{+0.8}_{-0.8}\times10^{-6}$ & $176^{+84}_{-49}$ & $1.3^{+0.9}_{-0.9}\times10^{-6}$ & $112^{+61}_{-41}$ & $1.3^{+1.0}_{-1.0}\times10^{-6}$ & $151^{+19}_{-64}$ & 113/92 & $7.63\times10^{-13}$\\ \hline
4XMM J174906.8 & $21^{+1}_{-1}$ & $1.3^{+0.1}_{-0.1}$ & $2.6^{+0.8}_{-0.6}\times10^{-3}$ & $6^{+4}_{-4}\times10^{-6}$ & $40^{+15}_{-15}$ & $3^{+40}_{-3}\times10^{-7}$ & $2^{+13}_{-2}$ & $3^{+44}_{-3}\times10^{-7}$ & $2^{+16}_{-2}$ & 400/442 & $1.12\times10^{-11}$\\ \hline
XMMU J175441.9 & $3^{+2}_{-1}$ & $0.3^{+0.4}_{-0.4}$ & $1.1^{+1.3}_{-0.6}\times10^{-5}$ & $2^{+1}_{-1}\times10^{-6}$ & $274^{+137}_{-137}$ & $3^{+14}_{-3}\times10^{-7}$ & $28^{+131}_{-28}$ & $1^{+1}_{-1}\times10^{-6}$ & $172^{+172}_{-172}$ & 33/26 & $4.31\times10^{-13}$\\ \hline
4XMM J175511.6 & $0.35^{+0.11}_{-0.09}$ & $1.1^{+0.1}_{-0.1}$ & $2.8^{+0.6}_{-0.5}\times10^{-5}$ & $1^{+6}_{-1}\times10^{-7}$ & $31^{+109}_{-31}$ & $6^{+7}_{-6}\times10^{-7}$ & $194^{+162}_{-188}$ & $<1\times10^{-6}$ & $<263$ & 69/76 & $3.06\times10^{-13}$\\ \hline
4XMM J175525.0 & $10^{+1}_{-1}$ & $1.4^{+0.2}_{-0.2}$ & $5^{+2}_{-2}\times10^{-4}$ & $<1\times10^{-6}$ & $<33$ & $9^{+16}_{-9}\times10^{-7}$ & $34^{+36}_{-34}$ & $<1\times10^{-6}$ & $<55$ & 171/170 & $1.84\times10^{-12}$\\ \hline
4XMM J175328.4 & $0.43^{+0.03}_{-0.02}$ & $1.40^{+0.04}_{-0.04}$ & $6.3^{+0.4}_{-0.3}\times10^{-4}$ & $3^{+2}_{-2}\times10^{-6}$ & $71^{+28}_{-29}$ & $2^{+2}_{-2}\times10^{-6}$ & $28^{+26}_{-28}$ & $4^{+3}_{-3}\times10^{-6}$ & $90^{+38}_{-36}$ & 883/914 & $5.56\times10^{-12}$\\ \hline
XMMU J180140.3 & $3^{+2}_{-1}$ & $2.1^{+1.3}_{-0.9}$ & $1.1^{+5.0}_{-0.8}\times10^{-4}$ & $2^{+3}_{-1}\times10^{-6}$ & $670^{+1005}_{-335}$ & $<4^{}_{}\times10^{-8}$ & $<8$ & $<6\times10^{-6}$ & $<1500$ & 25/27 & $2.31\times10^{-13}$\\ \hline

\end{tabular}
\tablefoot{The model used for spectral fit is power-law for the continuum plus three Gaussian at 6.4, 6.7, and 6.9 keV for iron emission complex. The spectral model is convolved with a Galactic absorption component \texttt{tbabs}. The normalization of power-law model $N_{\rm po}$ is given in units of photons keV$^{-1}$ cm$^2$ s$^{-1}$ at 1 keV and the normalization of the lines are given in units of photons cm$^2$ s$^{-1}$ in the line energy.}
\end{table}
\end{landscape}

\begin{figure*}
    \centering
    \includegraphics[width=\figsize\textwidth]{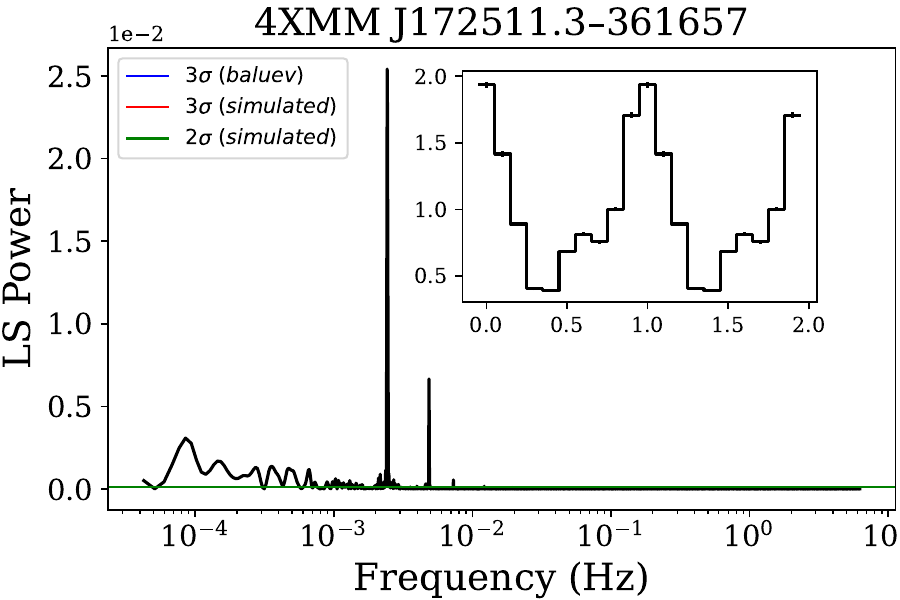}
    \includegraphics[width=\figsize\textwidth]{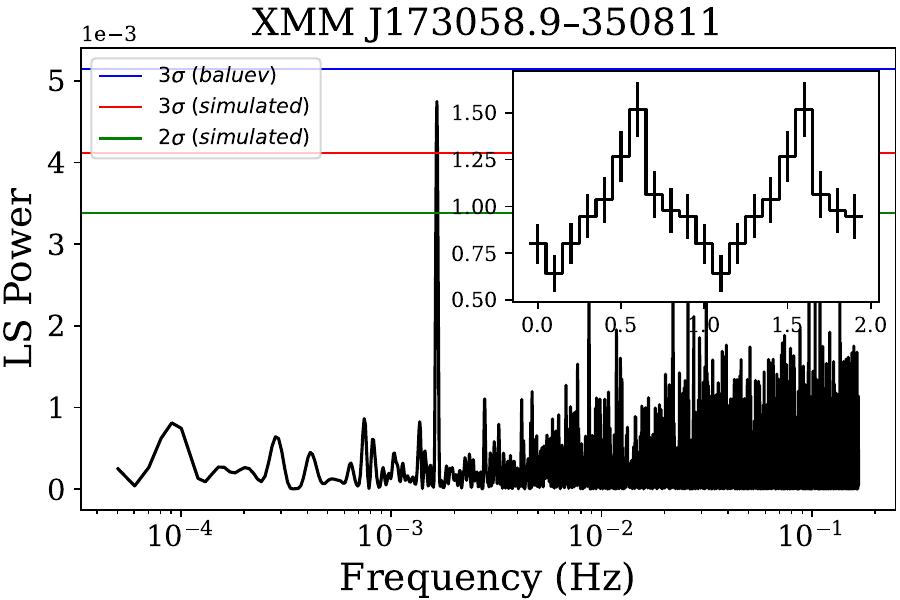}
    \includegraphics[width=\figsize\textwidth]{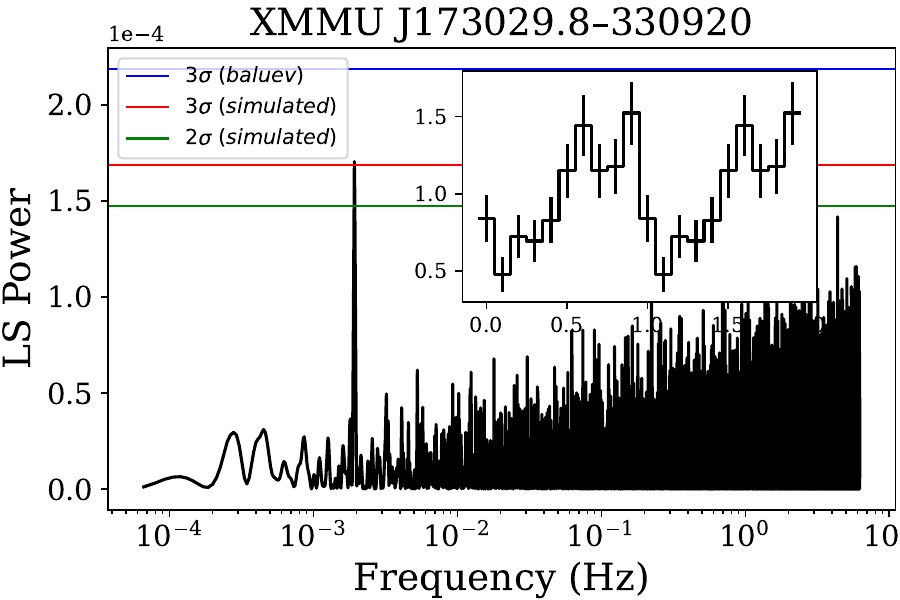}
    \includegraphics[width=\figsize\textwidth]{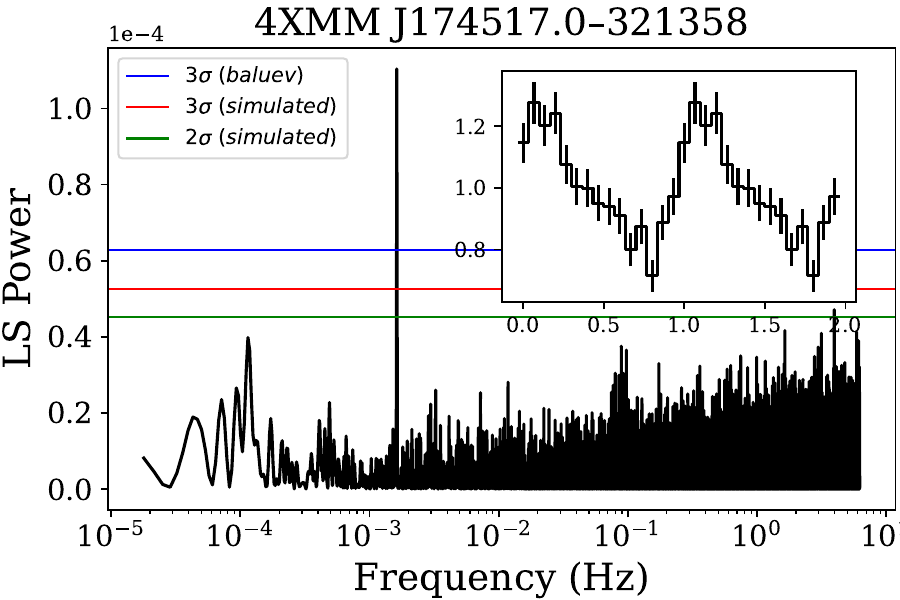}
    \includegraphics[width=\figsize\textwidth]{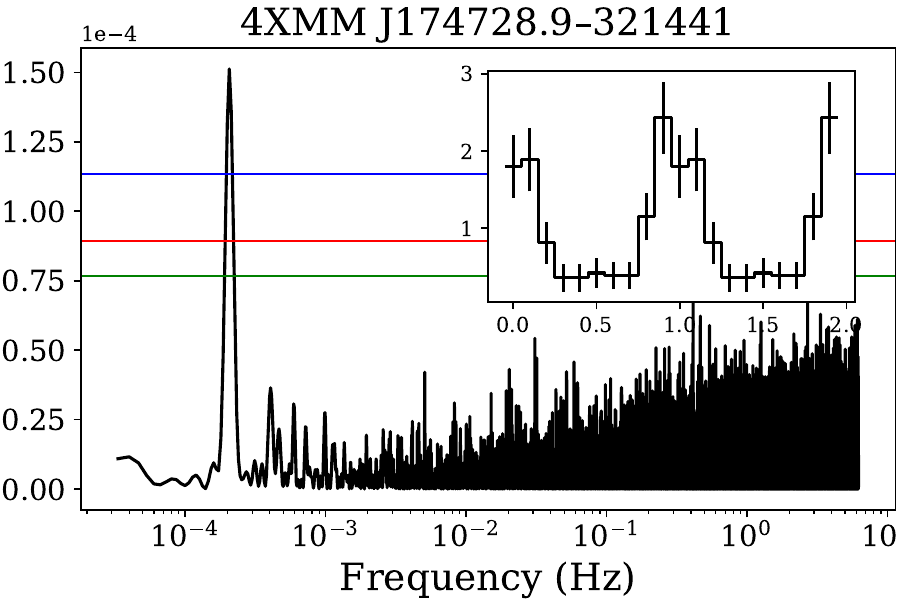}
    \includegraphics[width=\figsize\textwidth]{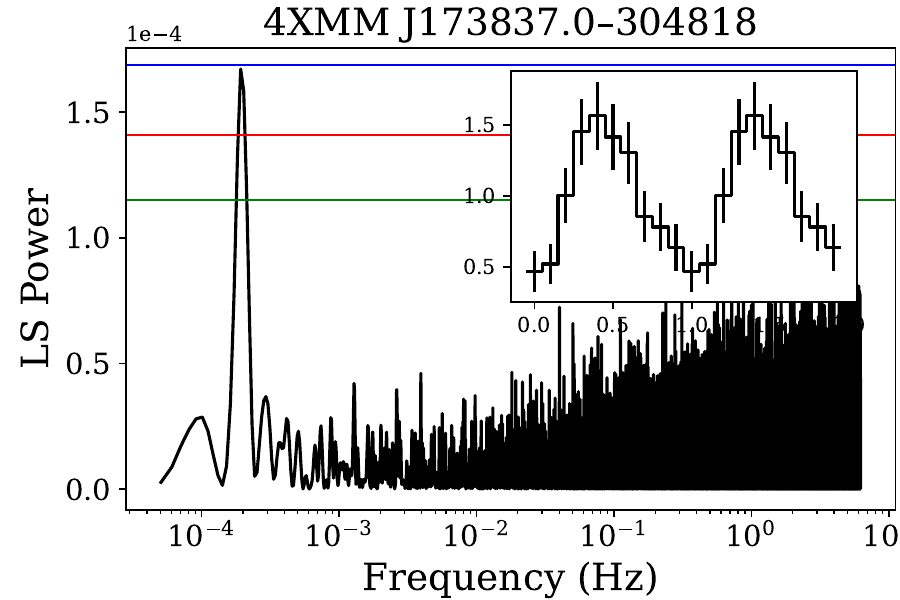}
    \includegraphics[width=\figsize\textwidth]{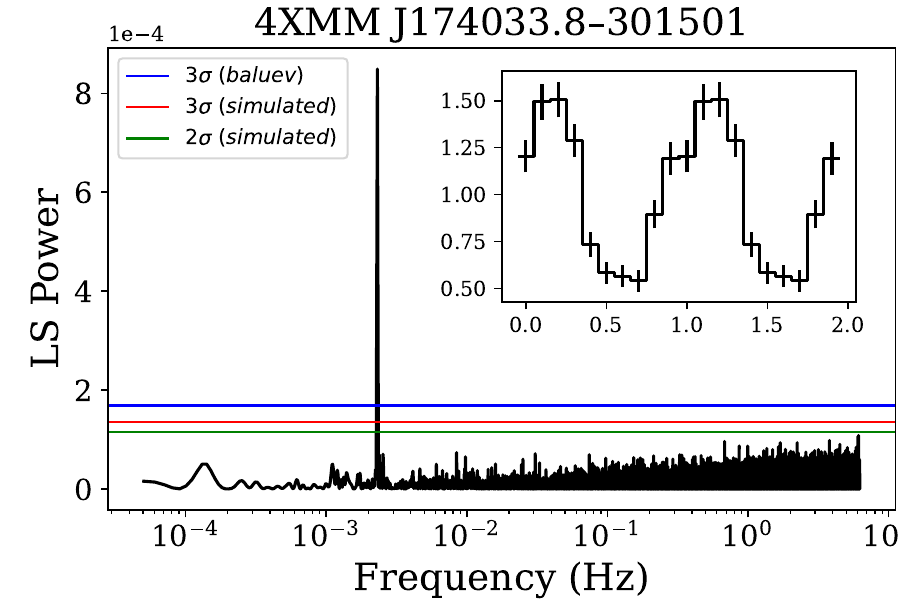}
    \includegraphics[width=\figsize\textwidth]{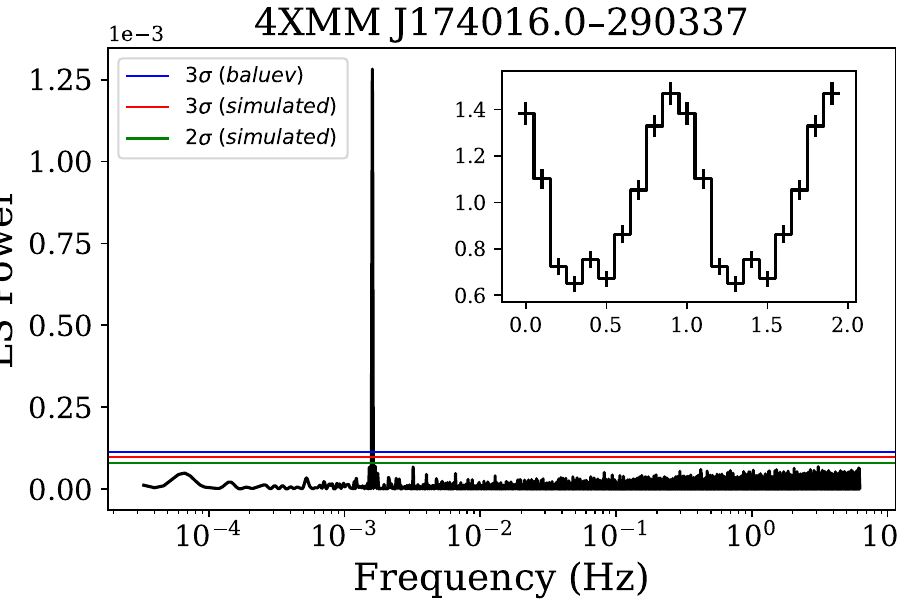}
    \includegraphics[width=\figsize\textwidth]{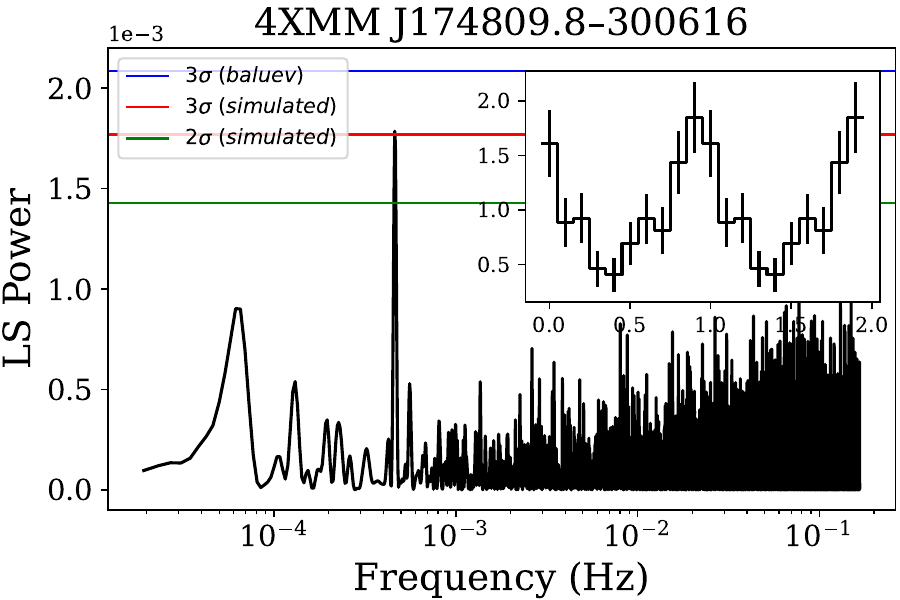}
    \includegraphics[width=\figsize\textwidth]{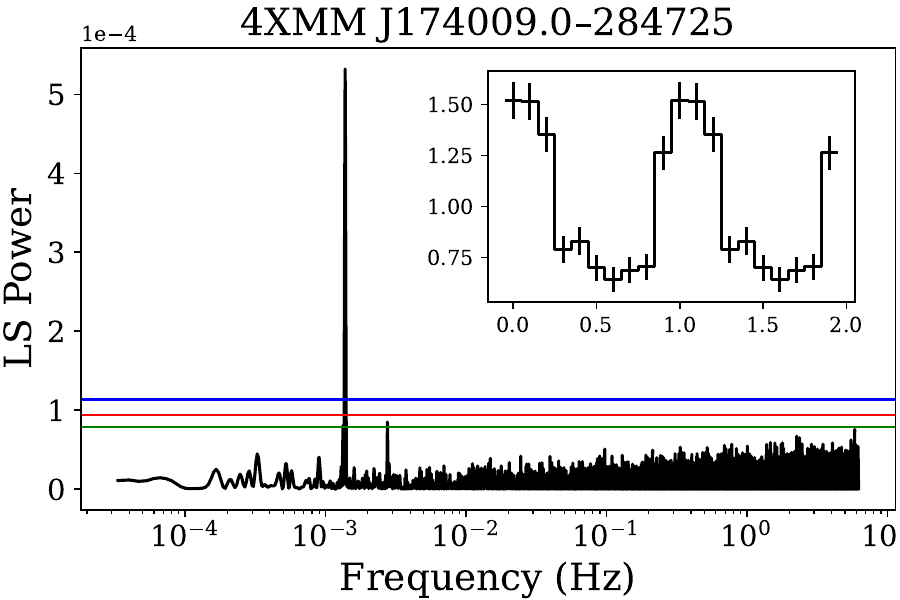}
    \includegraphics[width=\figsize\textwidth]{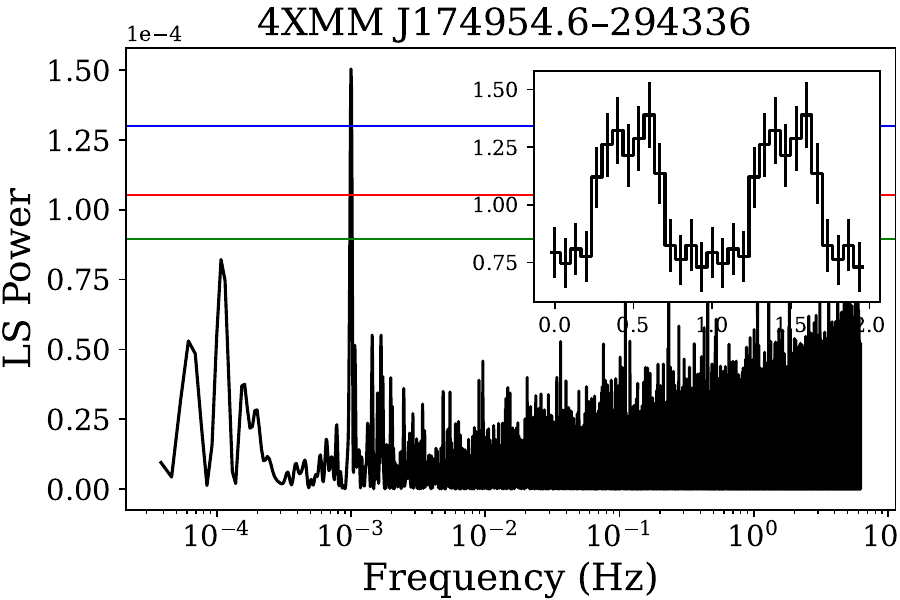}
    \includegraphics[width=\figsize\textwidth]{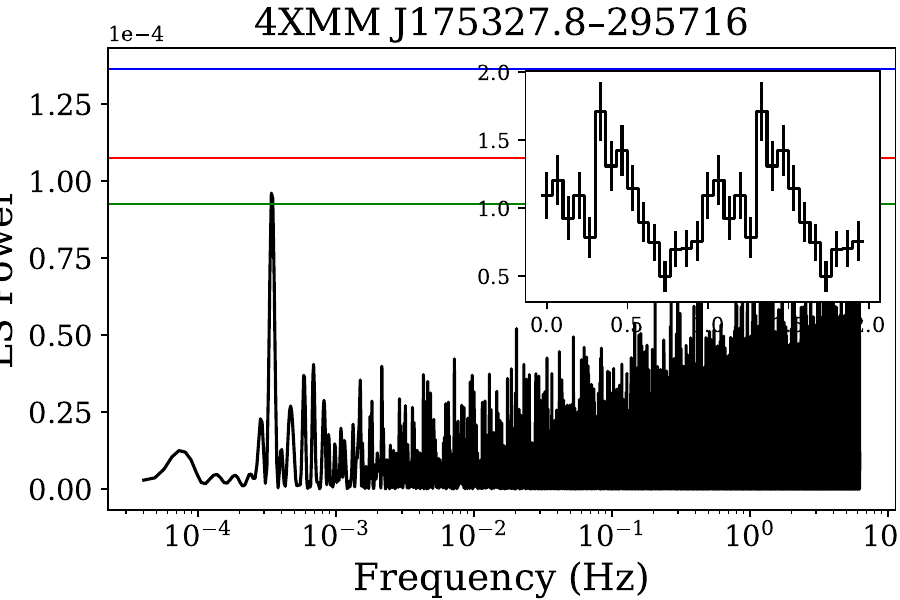}
    \includegraphics[width=\figsize\textwidth]{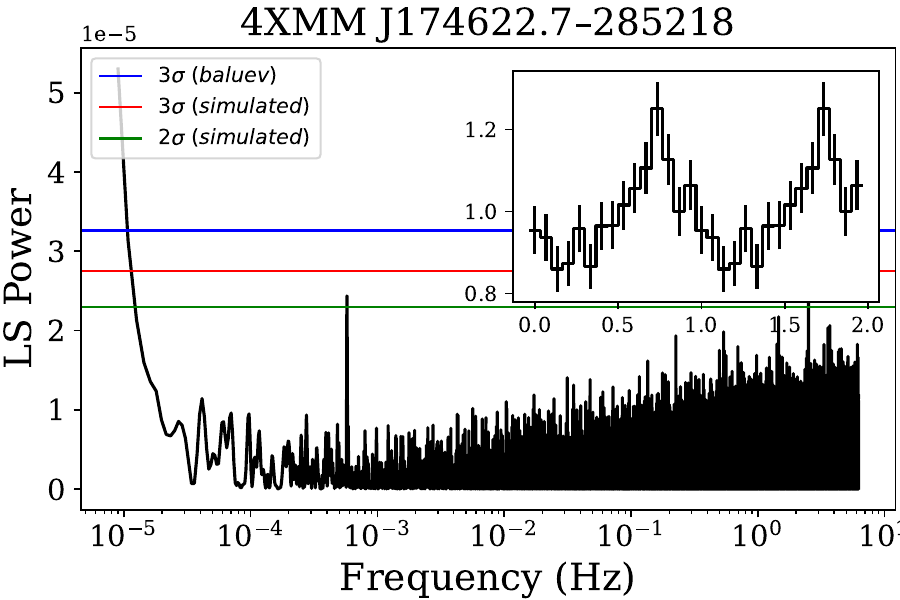}
    \includegraphics[width=\figsize\textwidth]{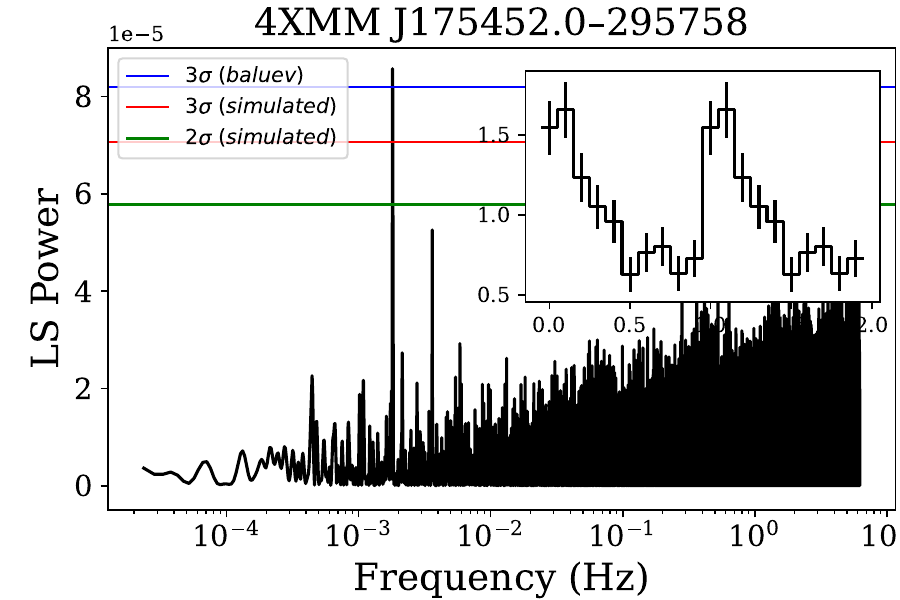}
    \includegraphics[width=\figsize\textwidth]{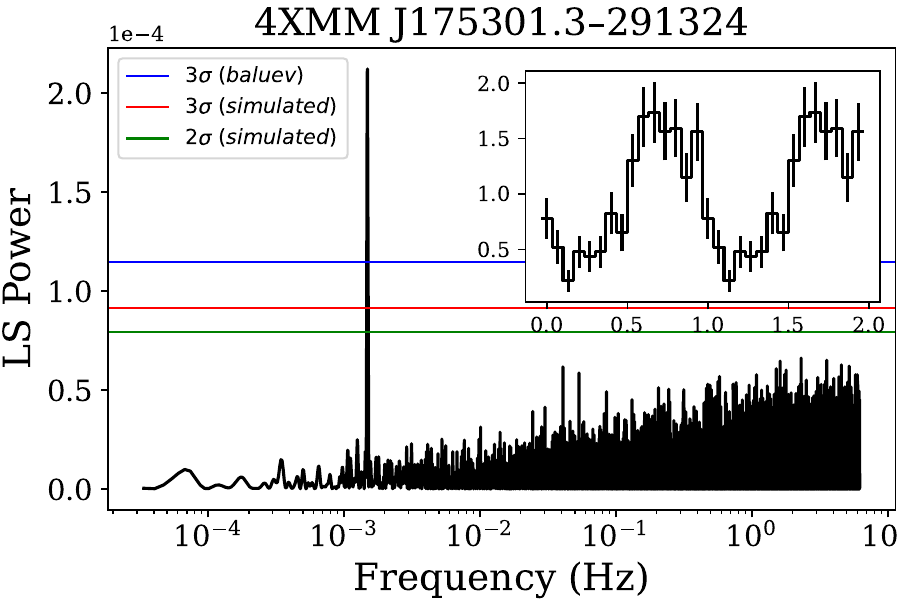}
    \includegraphics[width=\figsize\textwidth]{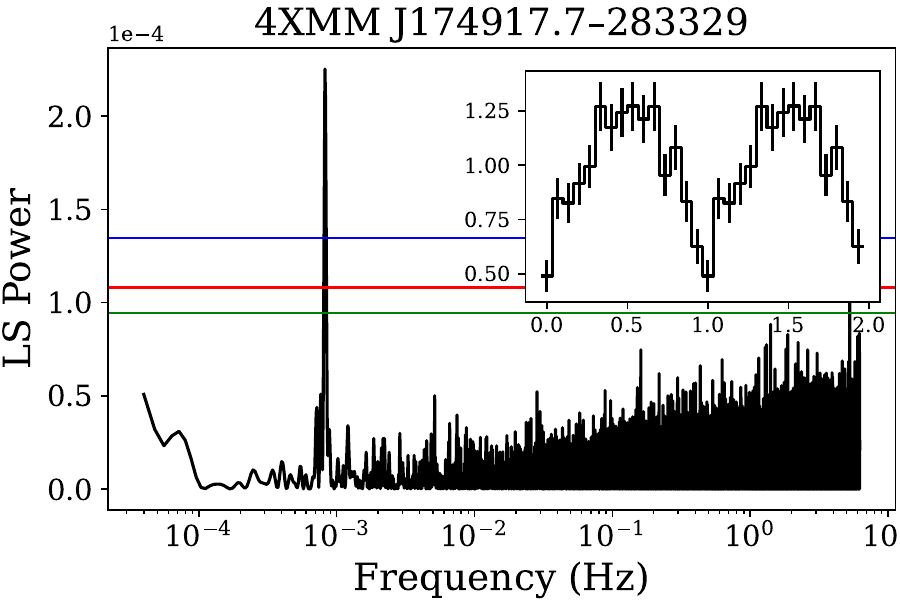}
    \includegraphics[width=\figsize\textwidth]{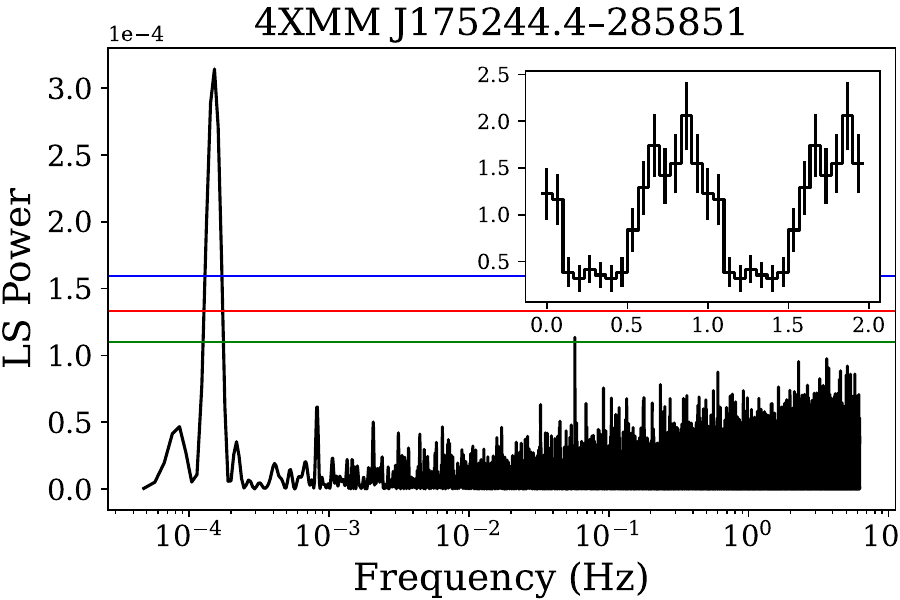}
    \includegraphics[width=\figsize\textwidth]{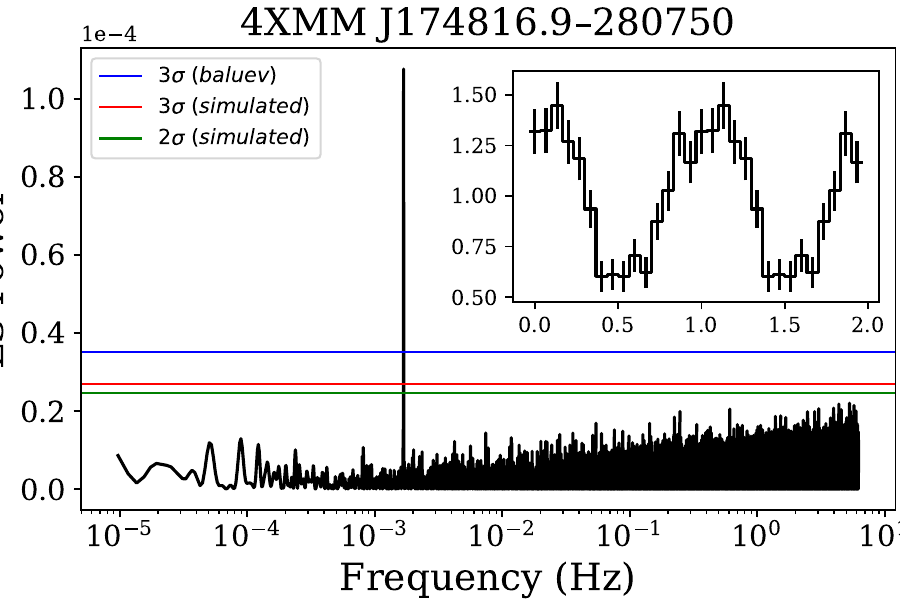}
    \caption{Lomb-Scargle periodogram of the sources listed in Table \ref{table:list_tab}. The periodograms are constructed using 2--10 keV EPIC-pn light curves. The horizontal green and red lines indicate the $2\sigma$ and $3\sigma$ confidence levels, respectively, computed from simulations, and the blue line indicates the false alarm probability ($3\sigma$ confidence level) estimated from the analytical approximation from \citet{baluev2008}. The small inset shows the folded light curve.}
    \label{fig:psd1}
\end{figure*}

\begin{figure*}
    \centering
    \includegraphics[width=\figsize\textwidth]{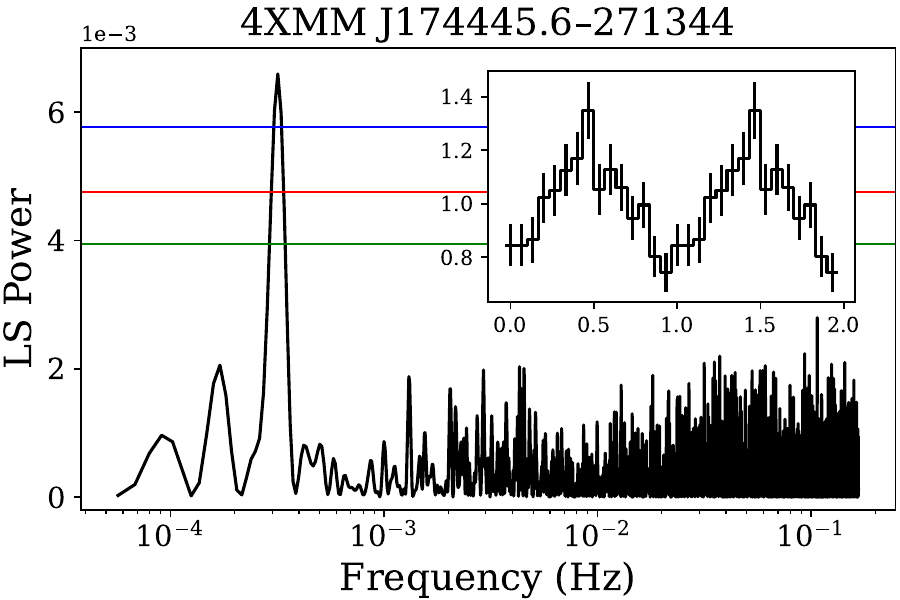}
    \includegraphics[width=\figsize\textwidth]{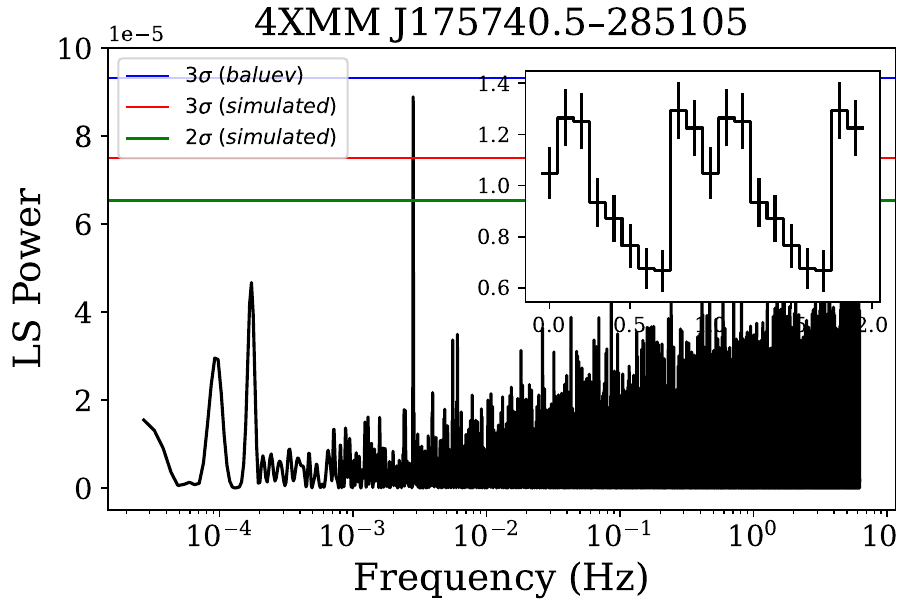}
    \includegraphics[width=\figsize\textwidth]{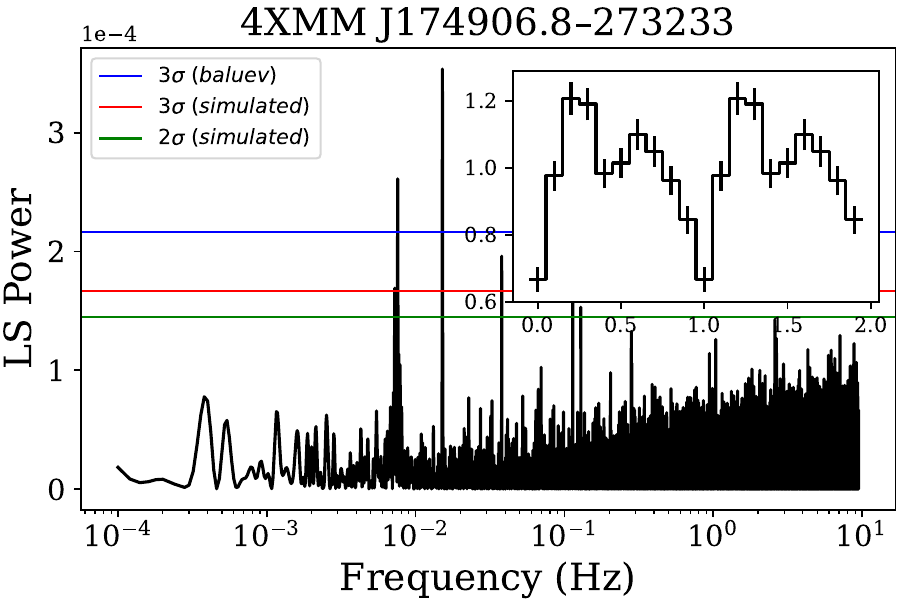}
    \includegraphics[width=\figsize\textwidth]{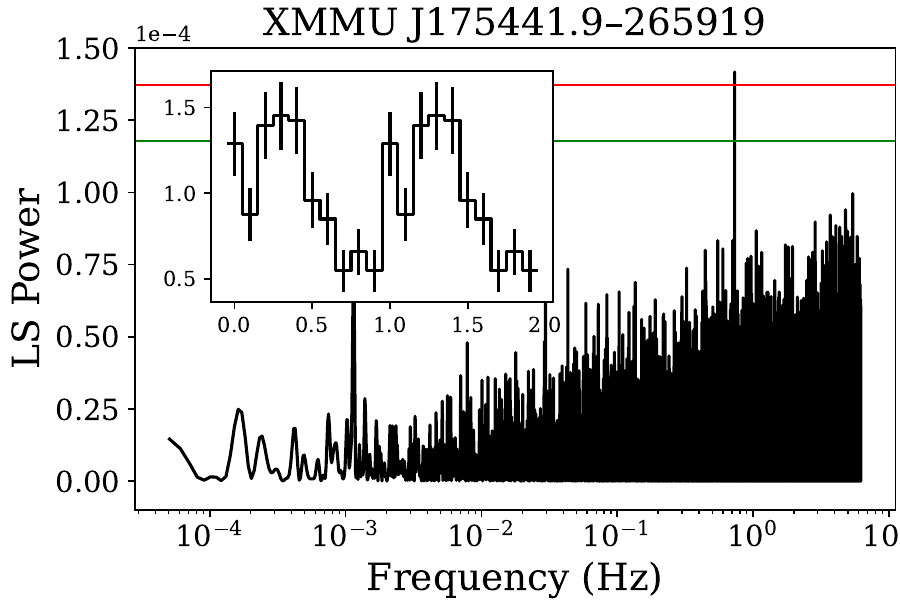}
    \includegraphics[width=\figsize\textwidth]{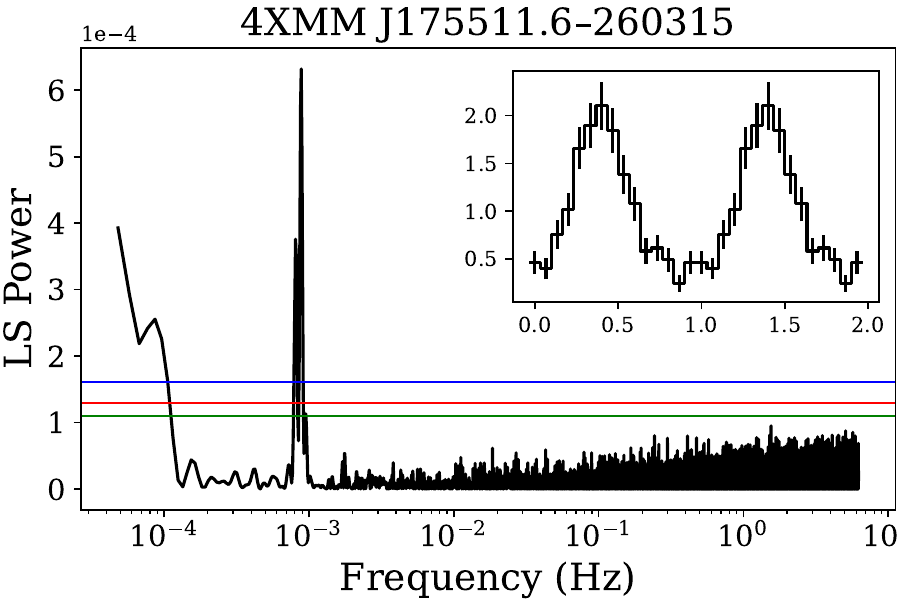}
    \includegraphics[width=\figsize\textwidth]{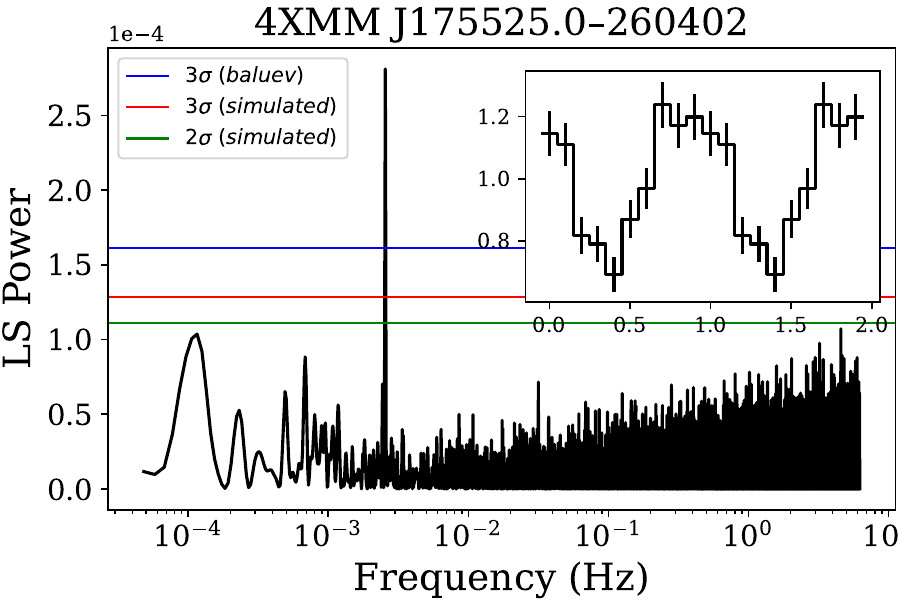}
    \includegraphics[width=\figsize\textwidth]{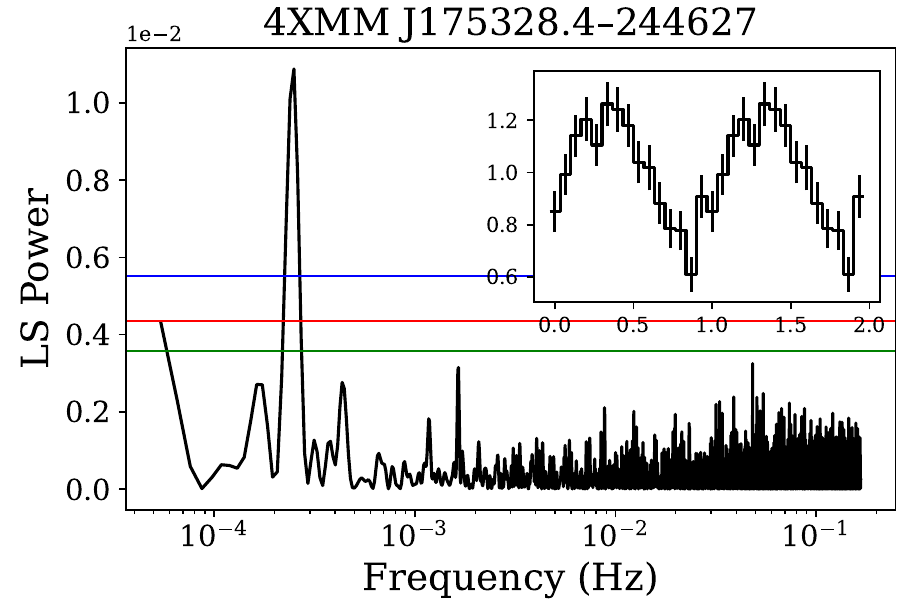}
    \includegraphics[width=\figsize\textwidth]{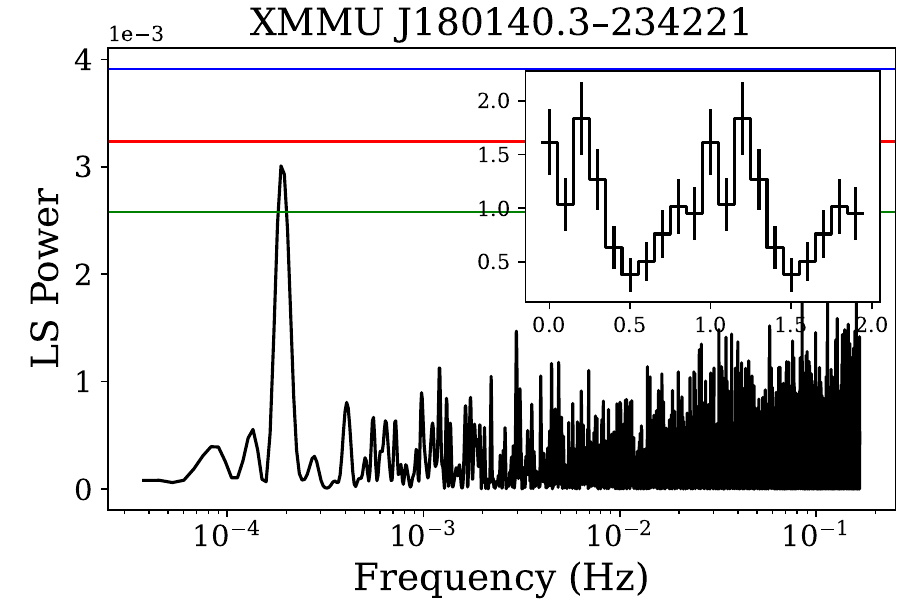}
    \caption{Fig. \ref{fig:psd1} Continued.}
    \label{fig:psd2}
\end{figure*}

\begin{figure*}
    \centering
    \includegraphics[width=\figsizee\textwidth]{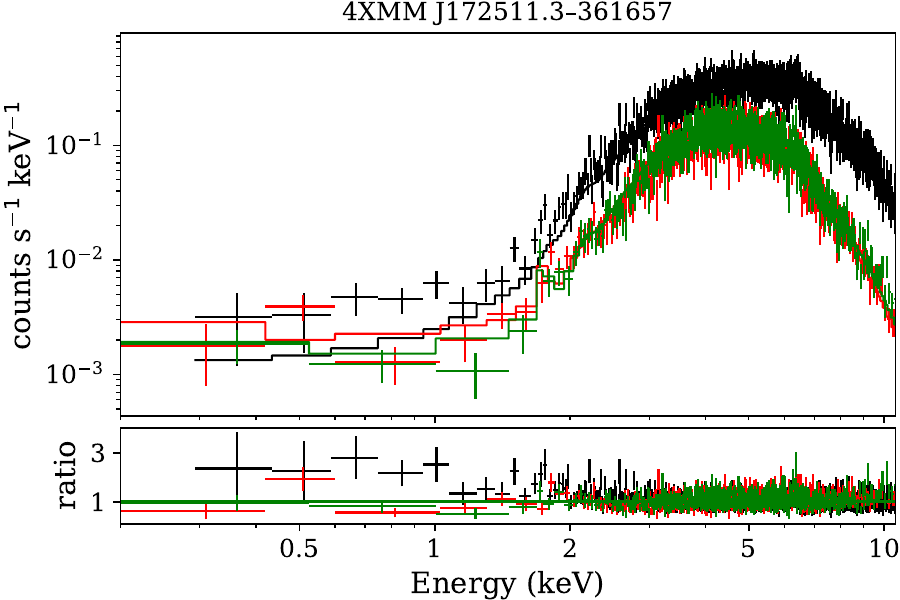}
    \includegraphics[width=\figsizee\textwidth]{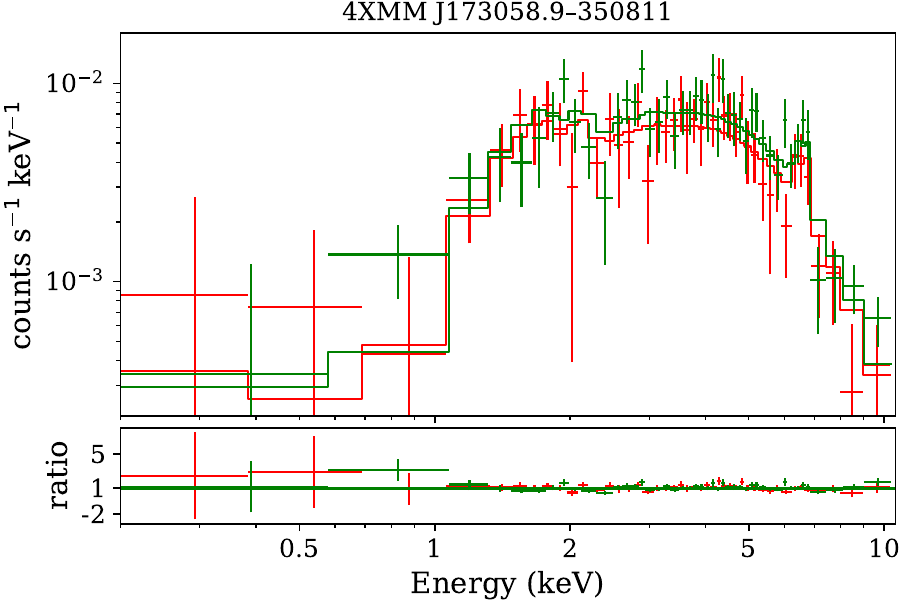}
    \includegraphics[width=\figsizee\textwidth]{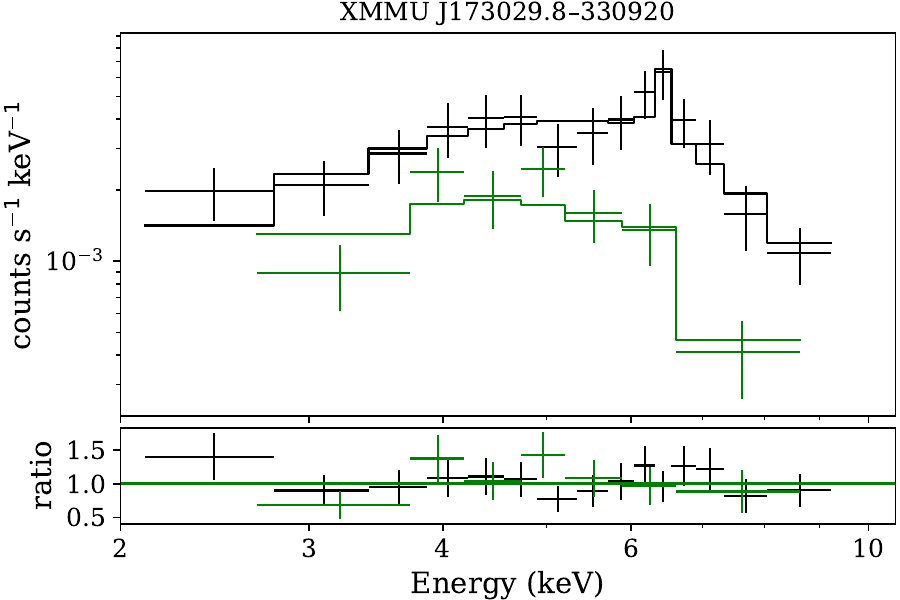}
    \includegraphics[width=\figsizee\textwidth]{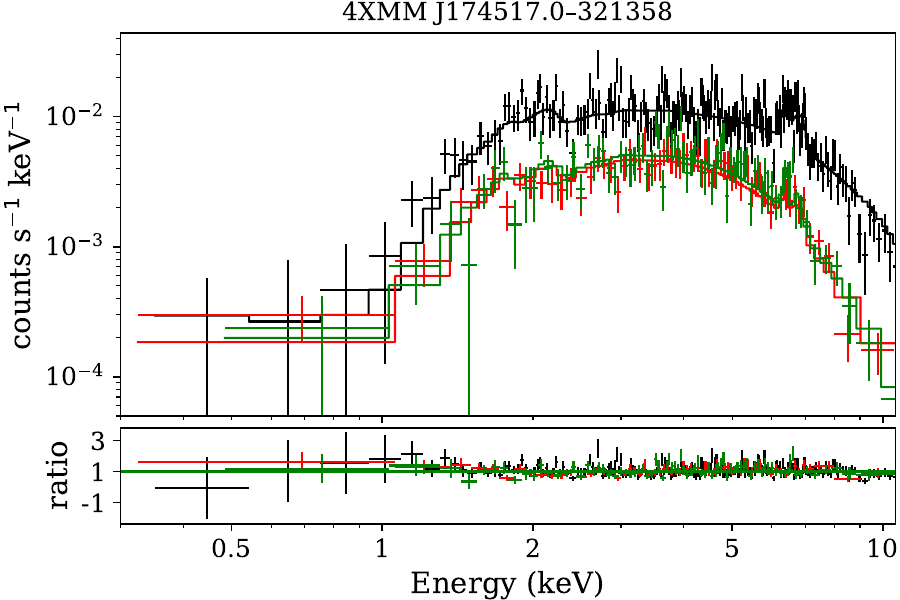}
    \includegraphics[width=\figsizee\textwidth]{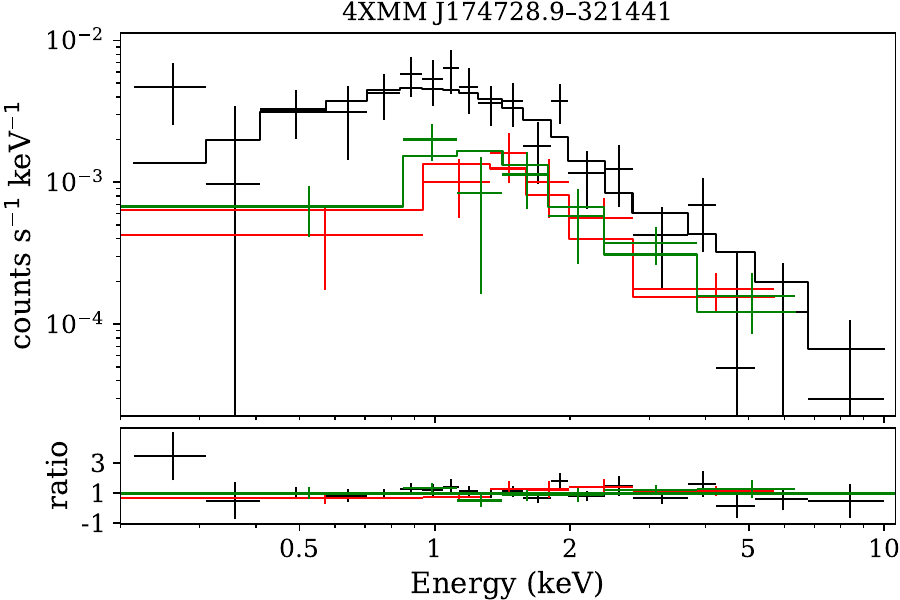}
    \includegraphics[width=\figsizee\textwidth]{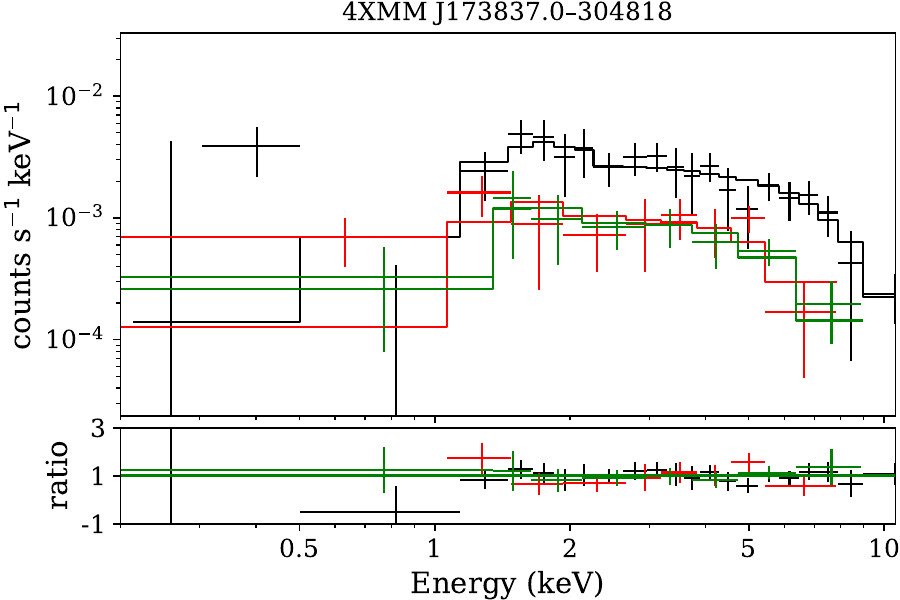}
    \includegraphics[width=\figsizee\textwidth]{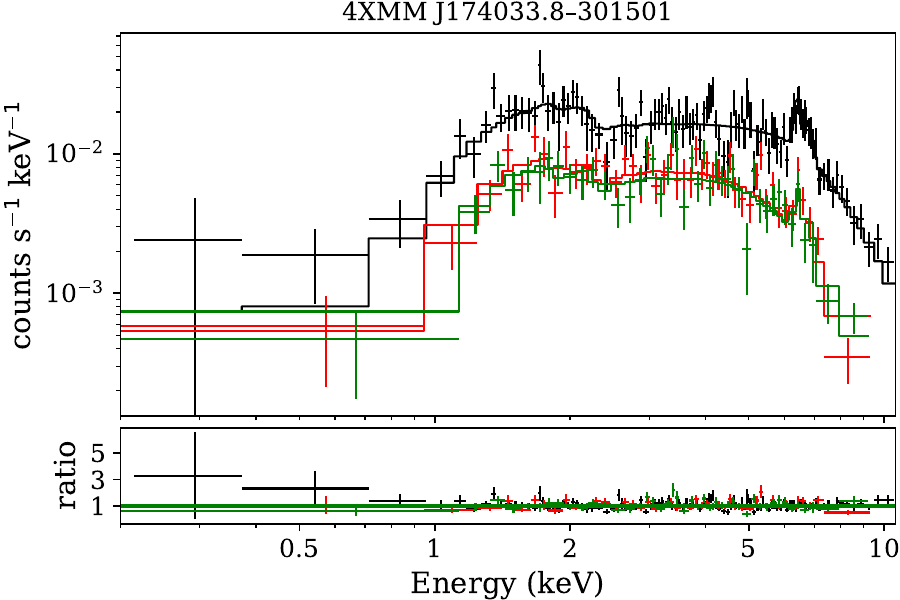}
    \includegraphics[width=\figsizee\textwidth]{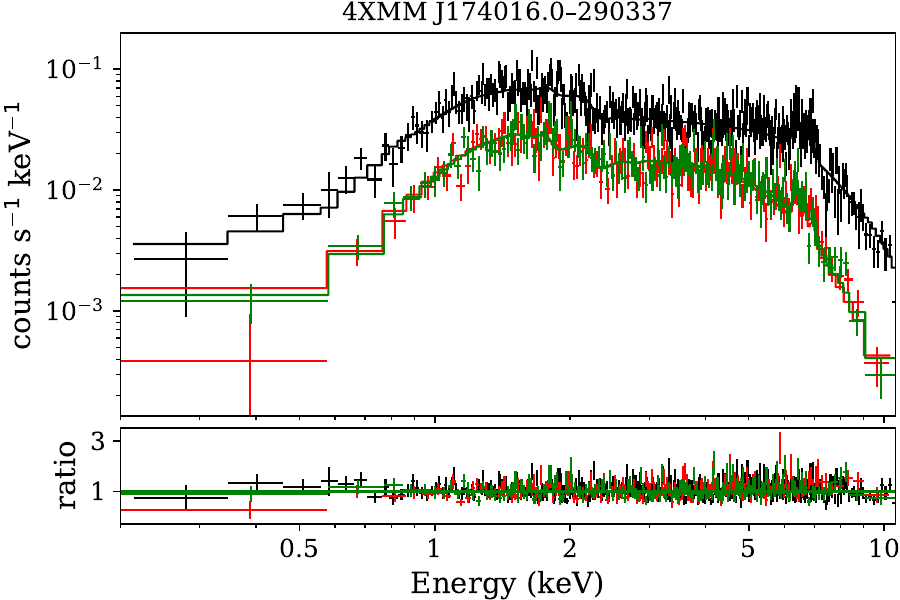}
    \includegraphics[width=\figsizee\textwidth]{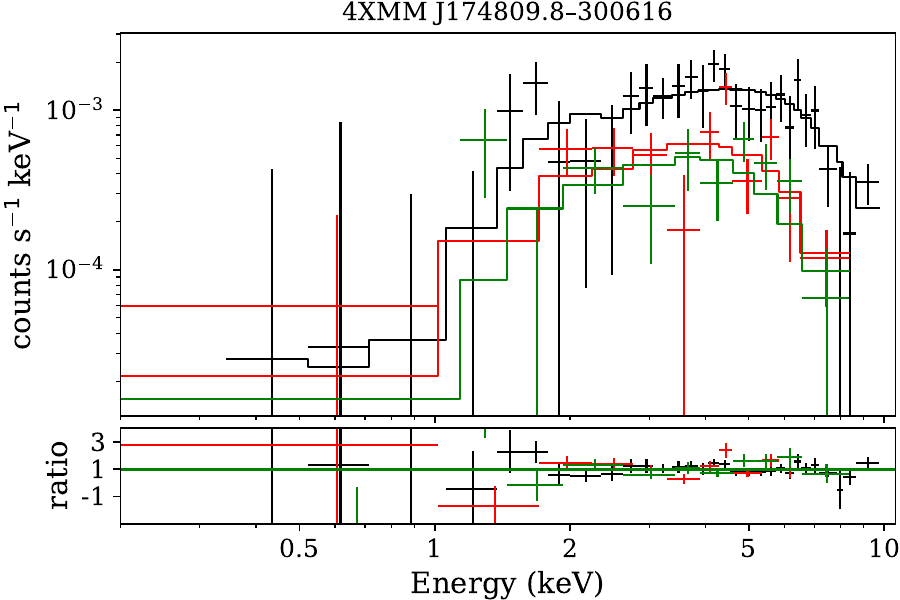}
    \includegraphics[width=\figsizee\textwidth]{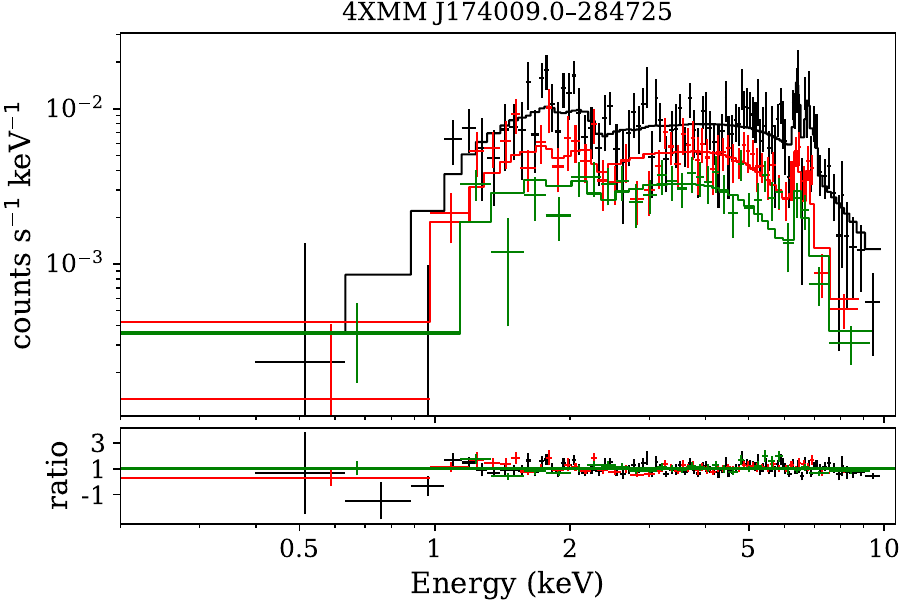}
    \includegraphics[width=\figsizee\textwidth]{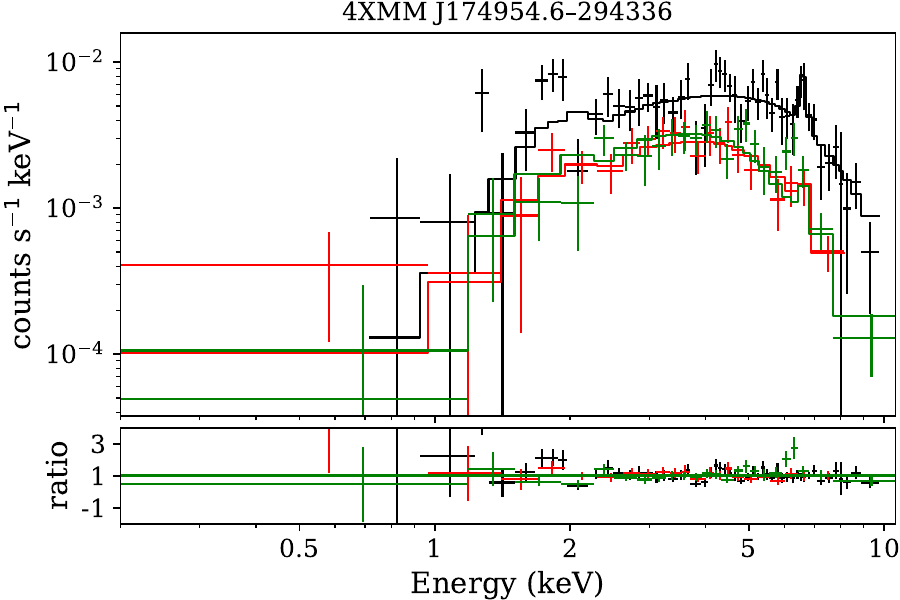}
    \includegraphics[width=\figsizee\textwidth]{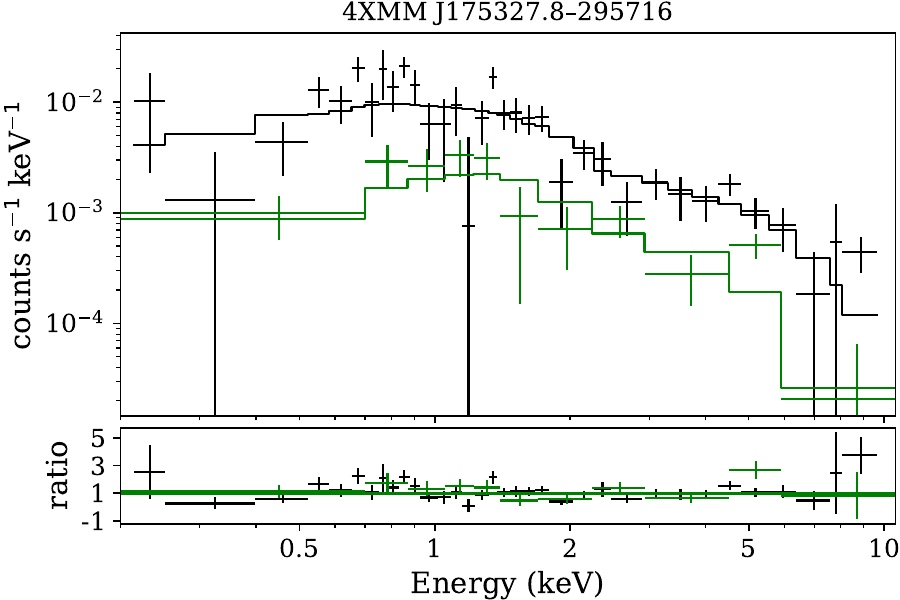}
    \includegraphics[width=\figsizee\textwidth]{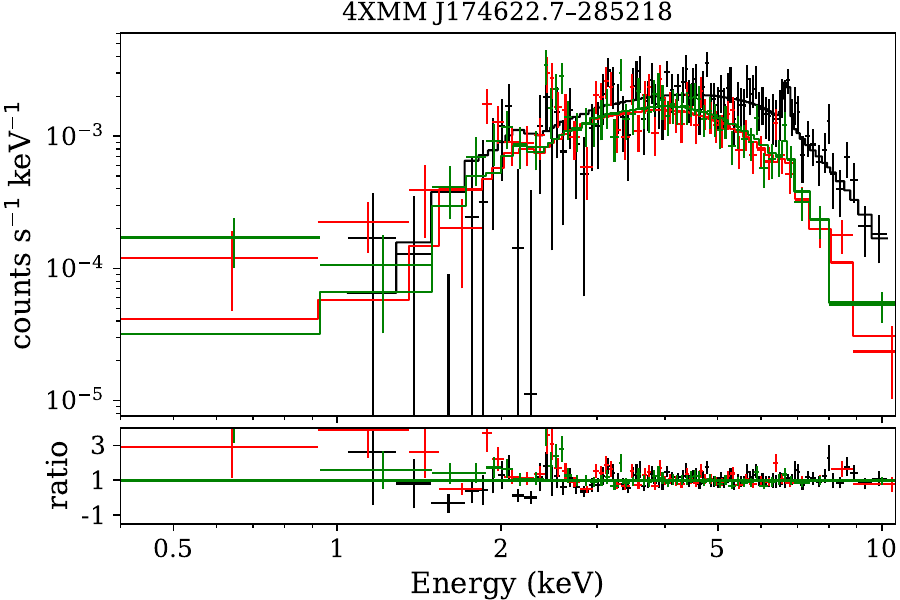}
    \includegraphics[width=\figsizee\textwidth]{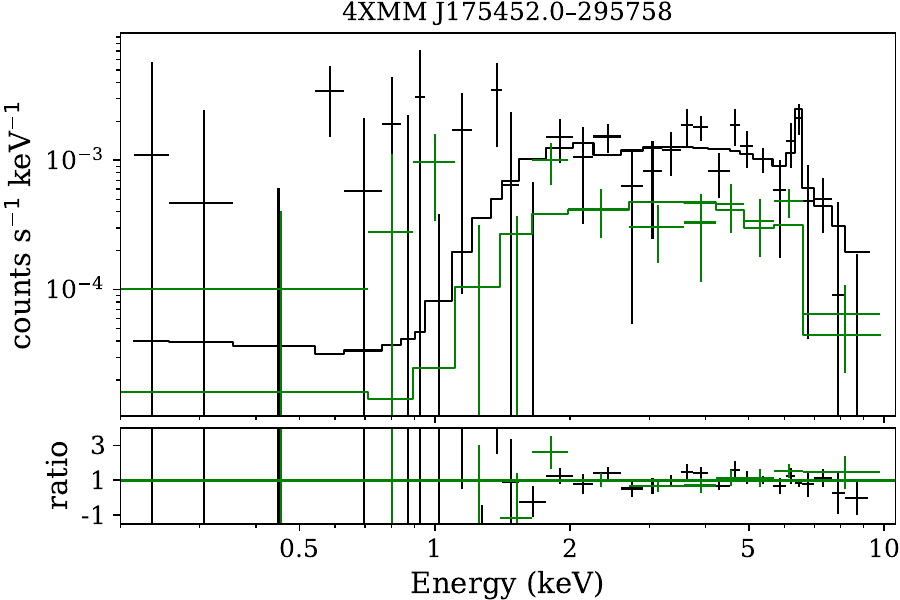}
    \includegraphics[width=\figsizee\textwidth]{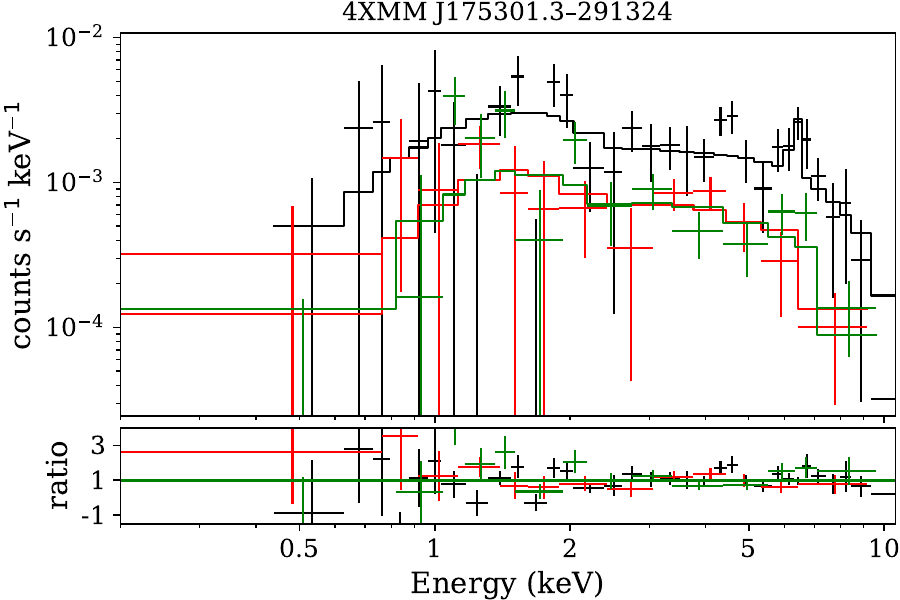}
    \includegraphics[width=\figsizee\textwidth]{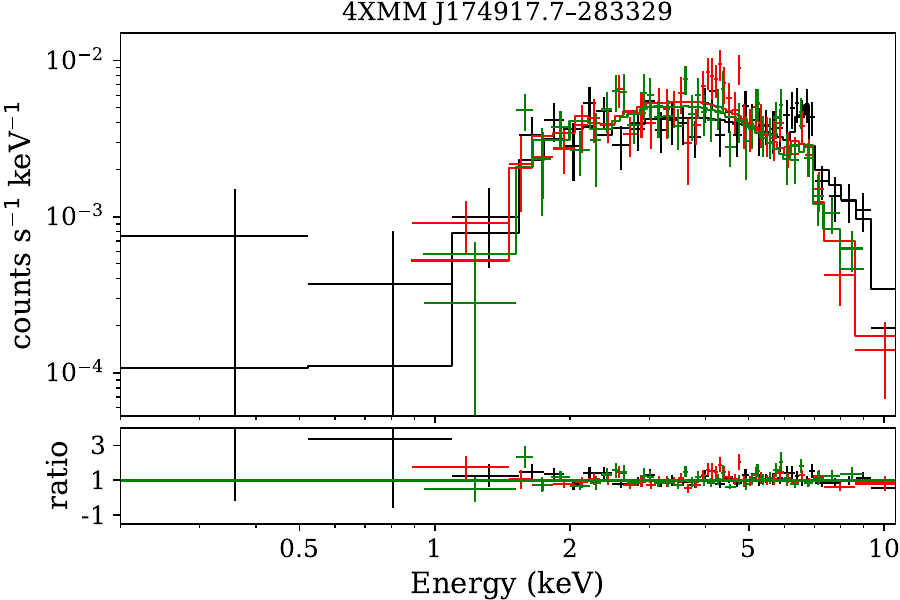}
    \includegraphics[width=\figsizee\textwidth]{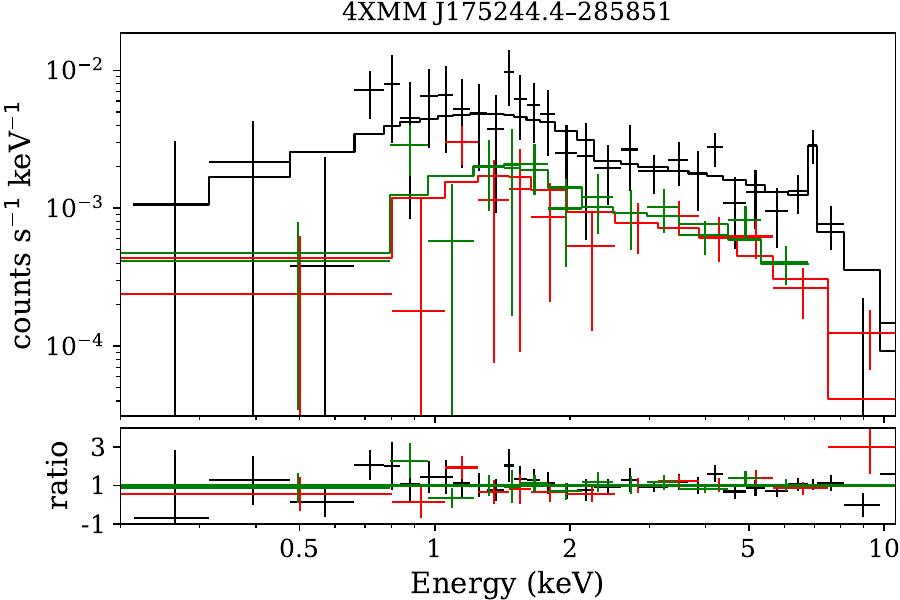}
    \includegraphics[width=\figsizee\textwidth]{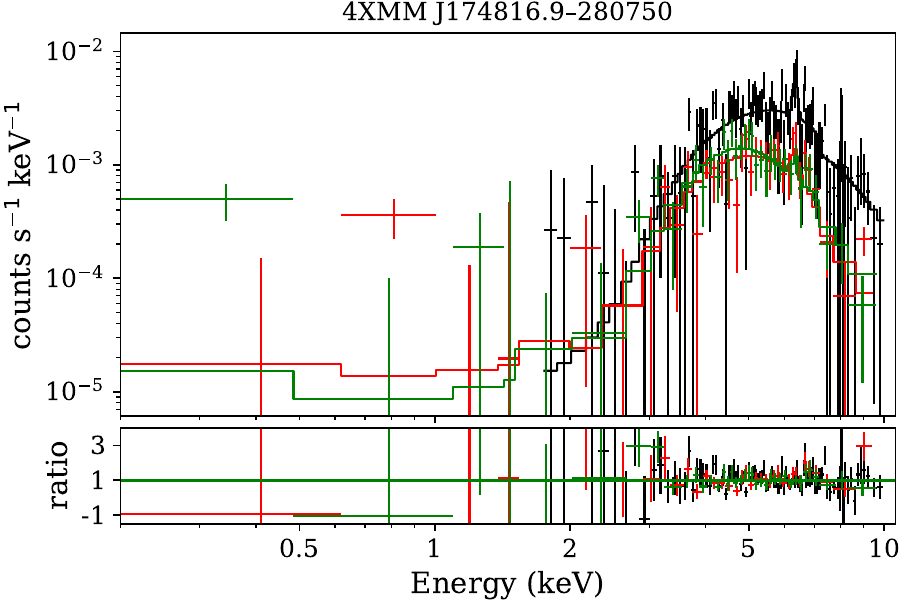}
    \caption{Spectral modeling of the sources in our sample using a model composed of \texttt{tbabs*(power-law+g1+g2+g3)}. The \texttt{g1},\texttt{g2}, and \texttt{g3} represent three Gaussian lines, at 6.4, 6.7, and 6.9 keV, respectively. The black, red, and green colors represent data from the EPIC-pn, MOS1, and MOS2 detectors, respectively.}
    \label{fig:spec1}
\end{figure*}

\begin{figure*}
    \centering
    \includegraphics[width=\figsizee\textwidth]{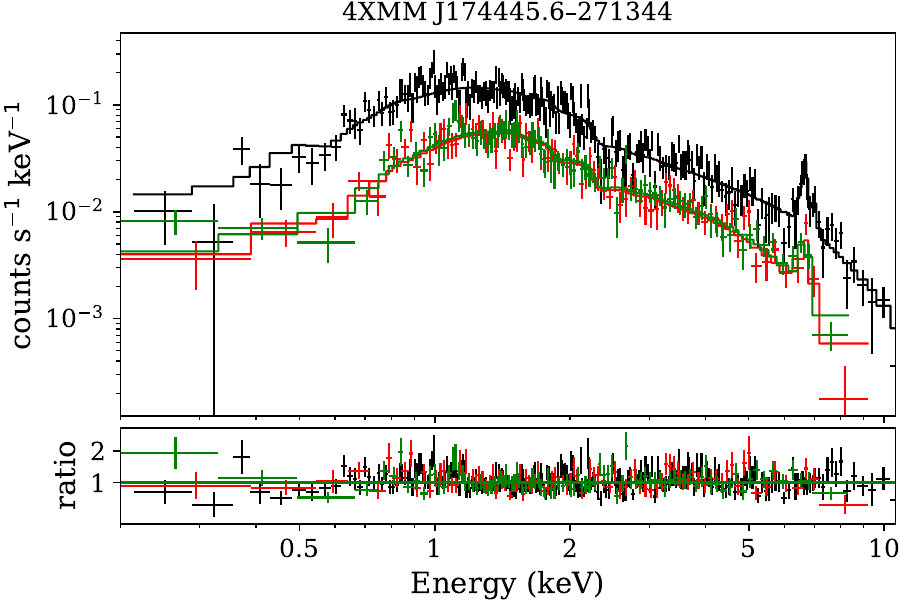}
    \includegraphics[width=\figsizee\textwidth]{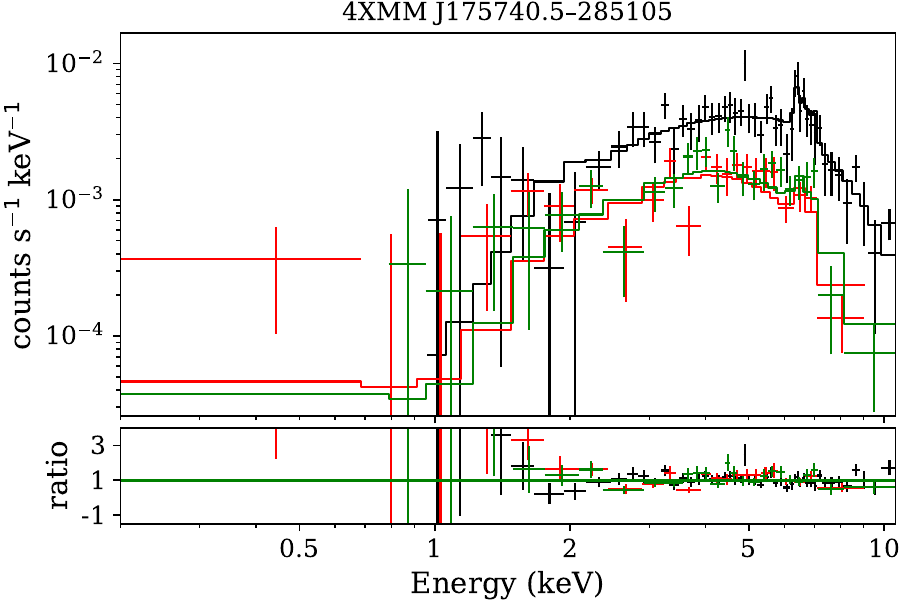}
    \includegraphics[width=\figsizee\textwidth]{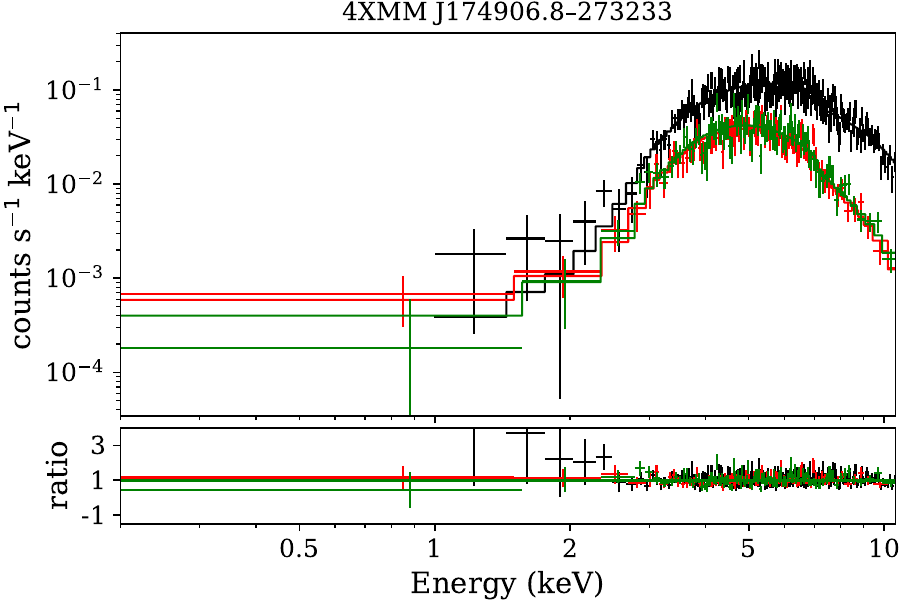}
    \includegraphics[width=\figsizee\textwidth]{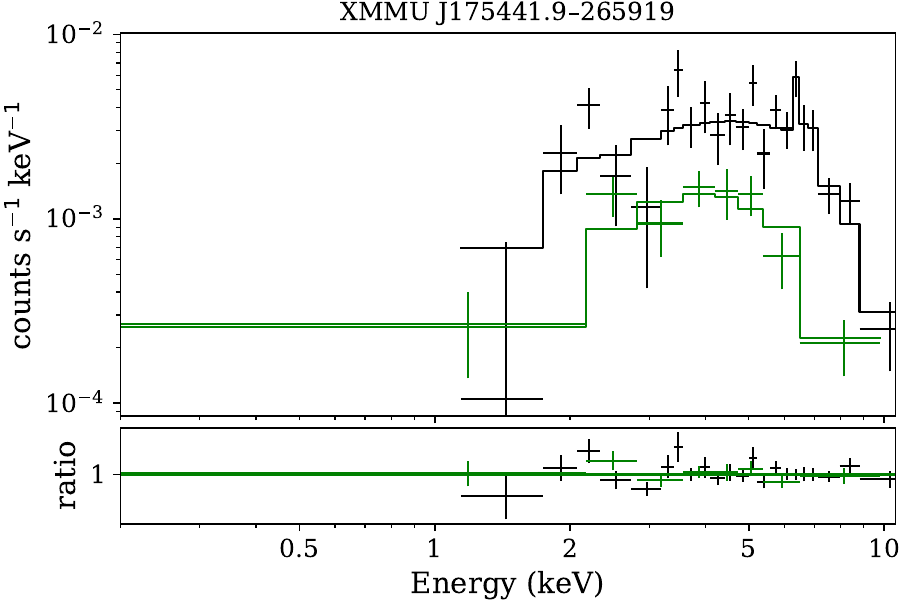}
    \includegraphics[width=\figsizee\textwidth]{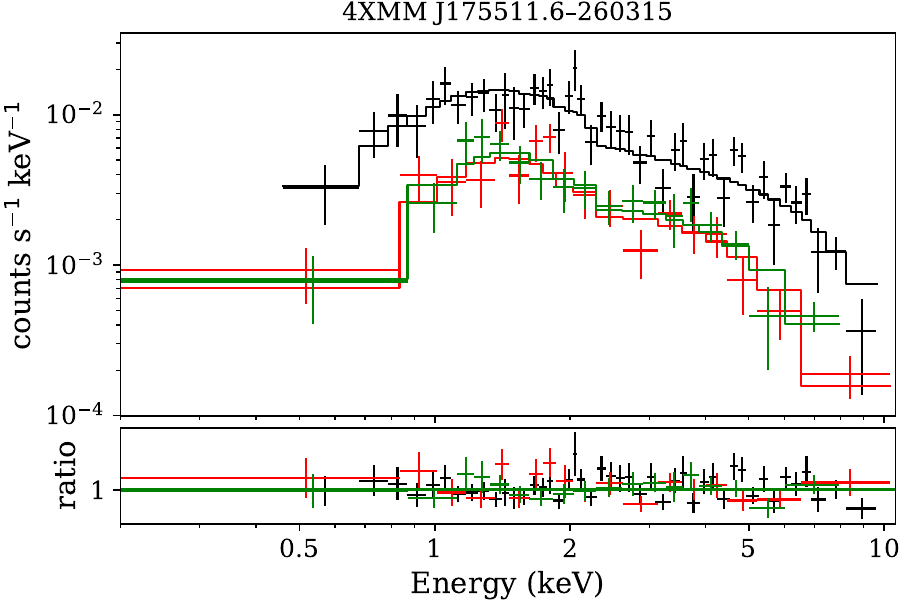}
    \includegraphics[width=\figsizee\textwidth]{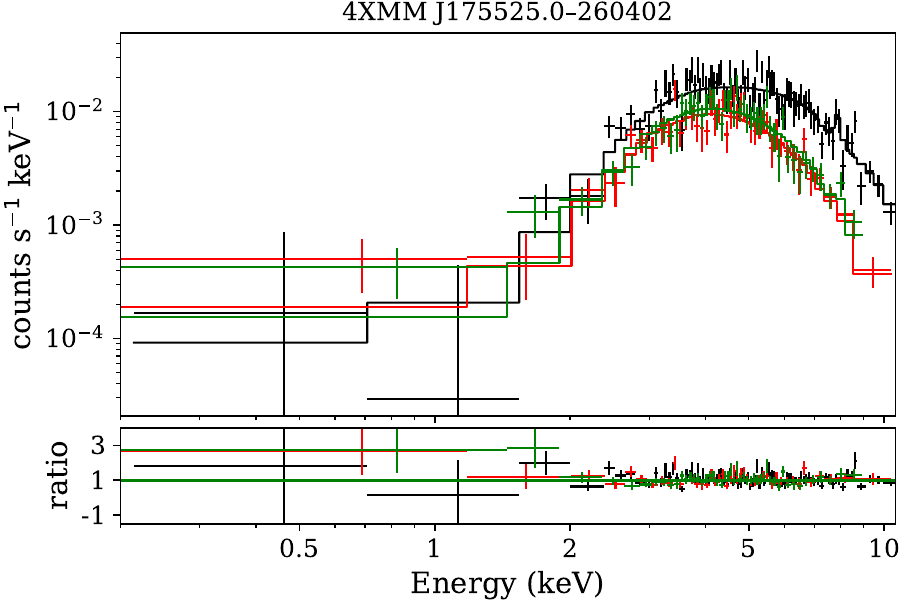}
    \includegraphics[width=\figsizee\textwidth]{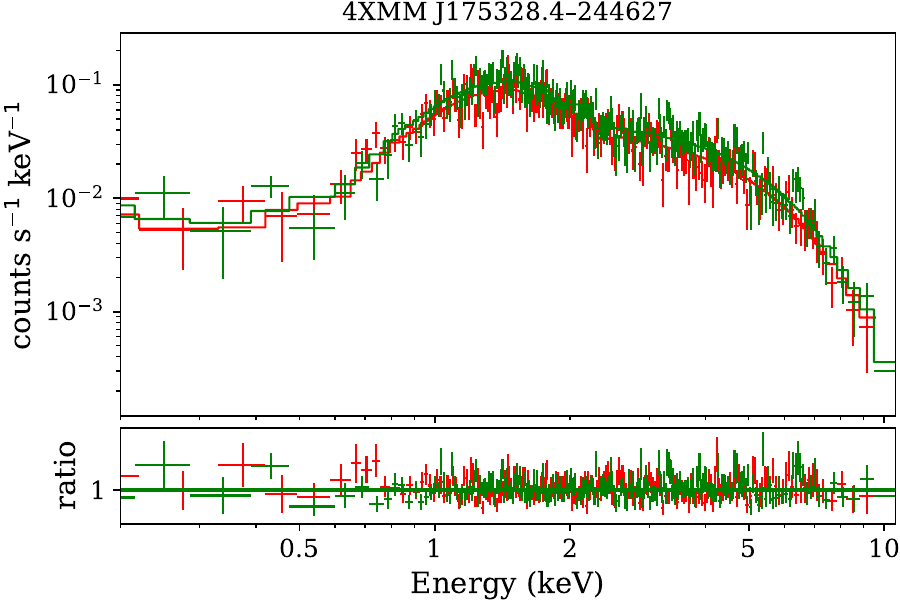}
    \includegraphics[width=\figsizee\textwidth]{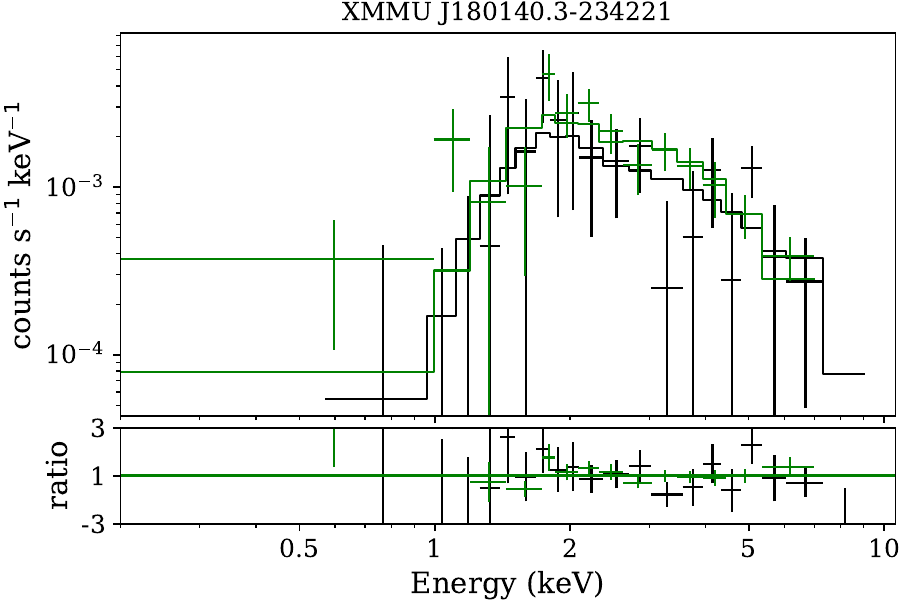}
    \caption{Fig. \ref{fig:spec1} Continued.}
    \label{fig:spec2}
\end{figure*}

\end{appendix}

\end{document}